\documentstyle[11pt]{article}

\newtheorem{theorem}{Theorem}[section]

\newtheorem{definition}[theorem]{Definition}
\title{Classical and Quantum Probability}
\author{R. F. Streater,\\Dept. of Mathematics,
King's College London,\\ Strand, WC2R 2LS}
\date{25 August 1999.}
\begin{document}
\maketitle
\begin{abstract}
We follow the development of probability theory from the beginning
of the last century, emphasising that quantum theory is really a
generalisation of
this theory. The great achievements of probability theory, such as
the theory of processes, generalised random fields, estimation
theory and information geometry, are reviewed. Their quantum versions
are then described.
\end{abstract}
\section{Introduction}
The are few mathematical topics that are as badly taught to physicists
as probability theory. Maxwell, Boltzmann and Gibbs were using probabilistic
methods long before the subject was properly established as mathematics.
Their language, of ensembles,
complexions, fluctuations and most probable state, are still used. When
quantum theory came along, the same notions were fitted into the new theory,
sometimes leading to confusion. We review the mathematical development
of probability, emphasising that quantum theory is a generalisation. The
approach to history is in the same spirit as used by Milligan in
\cite{Milligan}

There are three `philosophies' concerning probability. 
In the easy case, when there are finitely many possible outcomes to the
experiment being considered, 
Laplace's principle of equal ignorance tells us
that the probability of each of the outcomes is the same. In the case of a
die with six sides, experiments suggest that the probabilities are not all
exactly equal. Nevertheless, there is
not much error if we assume that the probability of each number is $1/6$.
An objection to Laplace's principle in general is that it is
not always clear that the outcome of a particular experiment is a matter
of chance, even when we do not know which outcome
will turn up; it could even be that a particular outcome is inevitable. Thus
a more robust version of Laplace's principle might be that in events
governed by chance, the probability of each possible outcome is the same.
This still leaves open the meaning of the phrase, `governed by chance'.
The difficulty of defining the `uniform' distribution when variates take
continuous values is illustrated by Bertrand's paradox (\cite{Decker},
p. 246). This demolished Laplace's principle for continuous variables.

The philosophy of Laplace applied to probability theory might be described as
{\em Platonic}. A real die is the shadow of the ideal die, which has
perfect sides and exact probabilities of $1/6$ for each outcome. This
has a modern form of expression: we model the real die by the sample space
$\Omega=\{1,2,\ldots,6\}$ whose elements $\omega$ are called {\em outcomes},
and assign the probability $1/6$ to each. The value of a {\em random
variable} $f$ is known if we know the outcome $\omega$; $f$ is
therefore a real-valued function on $\Omega$.
More generally, if the sample space is a finite set $\Omega$, an event $E$
is a subset of $\Omega$; we say that the event has occurred
if the $\omega$ that occurs lies in $E$. The probability that $E$ occurs
is the sum of the probabilities of the points in E:
\begin{equation}
p(E)=\sum_{\omega\in E}p(\omega).
\end{equation}
We say that two events, $E$, $F$, are independent if $p(E\cap F)=p(E)p(F)$.
In this way the binomial distribution can be derived for the total shown
by $n$ dice thrown independently, and all of Laplace's probability theory
can be derived. It can tell us what bets to lay on an event $E$, even
when only one trial is going to occur.

Laplace's method has been
successfully applied to statistical mechanics; the space of states is
discretised, thus avoiding Bertrand's paradox (the choice of
bins being suggested by quantum mechanics). Each bin is said to be
equally probable, and some hypotheses about independence is postulated.
Then it is shown that the complexion (macroscopic state) given by the
Gibbs distribution is not just the most probable, but is {\em
overwhelmingly} the most probable. The chance of any complexion minutely
different is put at $10^{-170}$. The Gibbs distribution is, of course,
the equilibrium state; if it is so probable, how come systems manage to
be out of equilibrium, and remain so for years at a time? This remark
is not aimed at Tolman \cite{Tolman}, who made it clear that the assumption
of equal probabilities applies only to equilibrium, and is to be tested
against experiment; it passes the test well, but he then spoils it by adding
as a further justification, `{\em without} this
postulate there would be nothing to correspond to the circumstance that
nature does not have any tendency to present us with systems in conditions 
which we regard as mechanically entirely possible but statistically
improbable'. The word `improbable' is itself based on Laplace's
assumption!

The second `philosophy' of probability can be described as Aristotelian;
it had taken hold by 1920, and is known as the
`frequentist' approach. It is essential that we can reproduce a long run
of independent trials each conducted under exactly the same experimental
conditions. In this respect, the theory makes sense only within a
scientific culture. Suppose that we have one
`variate', which may take continuous or discrete values. The result
of a measurement of the variate is assigned to one of a
preassigned set of `bins', which are intervals on the real axis.
We repeat a number of times, to find the histogram, that is, the number
$n_i$ of
events (out of N trials) in the $i^{\rm th}$ bin. If the histogram settles
down to a stable shape as we increase $N$, we declare that the value of the
variate is random (or, random enough). We then define the probability of
the event $i$ to be
\begin{equation}
p_i=\lim_{N\rightarrow\infty}\frac{n_i}{N}.
\end{equation}
This approach avoids the above problems that beset the Laplace philosophy.
However, it is completely useless as mathematics; a `definition'
should not depend on an infinite number of future experimental results.
There is not one theorem that can be
proved from this definition. Feller points out that we must avoid
confusion between a definition, and a method of measurement.
There is great heuristic value to the frequentist approach.
It is easy to teach \cite{Decker};
we do not prejudge the possible values that the variate
can have, or the probability of a given value; we can
introduce another variate $Y$, and observe its distribution, and its
joint probability distribution with $X$;
we can by extending this idea get access to the
joint probability distribution of any finite number of variates;
we can get some idea as to whether the variates
are random by examining a sequence of independent trials. We can even cover
situations in which two variates are not simultaneously observable, as in
quantum mechanics, by listing only the joint distributions of compatible
observables, and omitting those we cannot measure.
If we measure a variate $X$ with $n$ different values $x_i$ with relative
frequency $p_i$, we can construct a sample space $x_1,\ldots,x_n$, and
assign the probability $p_i$ to the occurrence of the outcome $x_i$.
Similarly, we can construct a sample space and probability for any
finite set of compatible variates if each measurement records their
values. The observed probabilities are more reliable than assuming all
points are equally probable.

However, there is one
grave disadvantage of the approach, apart from not being mathematics: it
is simply a description of data, and has much less predictive power than
Laplace's method. In particular, the method takes no position on the
question as to what are the possible variates. If
$\Omega$ has $|\Omega|=n$ points, then the random variables form a vector
space, denoted ${\cal A}(\Omega)$, of dimension $n$, so that at most $n$
random variables can
be linearly independent. No similar constraint holds in the frequentist
point of view. Thus a variate is not the same as a random variable.
In fact, it has no definition, other than the statement that its values are
random.

The frequentist approach is the safest one to use in studies
involving humans; social or financial matters are so complicated that
it is not likely that a sample space, $\Omega_1$ say, chosen to accommodate
the data observed so far, can describe all the possible new variates and the
values available to them. In the frequentist approach, faced with a new
variate, $Y$, one simply
takes the set of possible values of $Y$, say $\Omega_2=\{y_1,\ldots,y_m\}$,
and uses $\Omega_1\times\Omega_2$
as the sample space of the enhanced problem.

In a classical system in physics or chemistry, treated by classical
statistical mechanics, we want to follow the
scientific method: we model the system, do experiments, and
reject the model if forced to. In that case, we make another model,
estimate its parameters, and suggest more testing experiments.
We want and expect to be able to make predictions about variates not
measured yet. So we must reject the frequentist approach.

The third philosophy of probability \cite{Kolmogorov}was made clear by Kolmogorov, and
combines something of the first two;
it is to regard a probability theory as a {\em model}, to be tested
against experiment. It is like Plato's ideal, in that it
is based on a specified sample space $\Omega$;
but now the probability $p$ is not determined by
pure thought; any $p$ satisfying the axioms below provides us with a model.
\begin{definition}
Let $\Omega$ be a countable space.
A map $p:\Omega\rightarrow[0,1]$ is a {\em probability} if
$p(\omega)\geq0$ and $\sum_\omega p(\omega)=1$.
\label{probability}
\end{definition}
The probability of an event $E\subseteq\Omega$, and the concept of
independence of two events, are then as in Laplace's theory and clearly
depend on the choice of $p$.

A random variable $f:\Omega
\rightarrow{\bf R}$ is chosen to
represent the variate being observed, the particular choice being
part of the interpretation of the model. A theoretical idea, or else
the first few experiments on the variate $f$, allow us to get some
guide-lines for $\Omega$ and
the values of $p$. This is the subject of {\em estimation
theory}. We can judge the validity of the model (the choices we have made
for $(\Omega,p,f)$) by comparing the predictions of the model with the
observed frequencies $n_i$
using the theory of {\em significance tests}. Both estimation theory and
significance were developed before Kolmogorov's book. The
founders of these techniques were often frequentists; they realised
that one could not use an extreme frequentist
point of view: in estimation, they often postulated that the data had
Gaussian distributions, but with unknown parameters. In significance
testing, to make a start, they assumed a probability distribution
for the variate being measured; this is called the `hypothesis H',
which is part of the model; it can be rejected
if the data are significantly unlikely. This
has a version within Kolmogorov's formulation, in which we are given a
probability space, the pair $(p,\Omega)$, and model the variate with
a random variable, $f$.
To make contact with the well-established theory of estimation
and significance, we must relate the probability distribution of $f$
to the probability $p$. We now remind the reader how this is done.

Given a finite probability space $(p,\Omega)$ and a
random variable $f:\Omega\rightarrow {\bf R}$, the {\em probability
distribution} of $f$ is denoted $p_f(i)$, and is determined as follows:
let $x_i, i=1,2,\ldots,n$ be the values that
$f$ takes, and let $p_f(i)$ be the probability that the event $\{\omega:f
(\omega)=x_i\}$. That is
\begin{equation}
p_f(i):=\sum_{\omega:f(\omega)=x_i}p(\omega).
\end{equation}
This is what is accessible to experiments when we measure $f$.
The mean of $f$ is determined by $x_i$ and $p_f$:
\begin{equation}
E_p[f]:=\sum_\omega p(\omega)f(\omega)=\sum_i x_ip_f(i),
\mbox{ also written }p.f.
\label{mean}
\end{equation}
Given two random variables on $(\Omega,p)$, $f,g$ we define the joint
distribution, denoted $p_{f,g}(i,j)$ to be
\begin{equation}
p_{f,g}(i,j):=p\{\omega:f(\omega)=x_i\mbox{ and }g(\omega)=y_j\}.
\end{equation}
We say two r. v. are independent if the events $\{f(\omega)=x_i\}$ and
$\{g(\omega)=y_j\}$ are independent for all $i,j$. This is equivalent to
the frequentists' version: $p_{f,g}(i,j)=p_f(i)p_g(j)$.
The joint distribution determines $p_f$ and $p_g$ as its marginals,
and also all moments, e. g. the cross-moment $E_p[fg]$
can be shown to be $\sum_{ij}x_iy_jp_{f,g}(i,j).$

A probability $p$ defines a linear functional on the set ${\cal A}(\Omega)$
by the expectation, (\ref{mean}): $f\mapsto E_p[f]$.
We shall call any such functional a {\em state}: it is linear and positive,
taking the value $1$ on the sure function $I$.
The dual space ${\cal A}^d$ is the set of all linear functionals, so the
states form a subset of ${\cal A}^d$; it is denoted $\Sigma(\Omega)$.
Given two states $p_1$ and $p_2$ and $0<\lambda<1$, their mixture
with probabilities $\lambda$ and $1-\lambda$, $p=\lambda p_1+(1-
\lambda)p_2$, is again a state. So the states form a convex set.

Whether $\Omega$ is countable or not, for a random variable $f$ on
$(\Omega,p)$
the probability of the occurrence of a single value $f_0$ might be zero,
even when there is an $\omega_0\in\Omega$ with $f(\omega_0)=f_0$;
for, $p(\omega_0)$ might be zero. This often happens when $\Omega$ is
not countable, and $f$ takes continuous values. Then, more information
about the probability measure is provided by the `cumulative'
distribution function
\begin{equation}
P_f(x)=p\{\omega:f(\omega)<x\}.
\end{equation}
This is an increasing function of $x$, going from $0$ at $x=-\infty$ to $1$
at $x=\infty$.
We say that $f$ possesses a density $\rho_f$ if $P_f(x)$ is
differentiable, and we write
\begin{equation}
\rho_f(x)=\frac{dP_f(x)}{dx}.
\end{equation}
It is clear that we cannot cope with this subject without a certain
amount of real analysis.

A cumulative probability distribution $P_f(x)$ is determined by its
characteristic function
\begin{equation}
C_f(\lambda):=\int e^{i\lambda x}dP_f(x)
\end{equation}
Here we use the Stieltjes integral. Any characteristic function satisfies
\begin{enumerate}
\item $C(\lambda)$ is continuous;
\item $C(0)=1$;
\item $C$ is of positive type:
\[\sum_{ij}\overline{z}_iz_jC(\lambda_j-\lambda_i)\geq0.\]
\end{enumerate}
Conversely, any function $C$ obeying $1,2$ and $3$ is the characteristic
function of a probability distribution; this is Bochner's theorem.
In terms of the original $(\Omega,p)$ and random variable $f$, the
characteristic function is
\begin{equation}
C_f(\lambda):=E_p[e^{i\lambda f}].
\end{equation}
If $C_f$ is analytic in $\lambda$ around $\lambda=0$, we can easily
justify the expansion
\[C_f(\lambda)=\sum_n(i\lambda)^nE_p[f^n]/n!=\sum_n(i\lambda)^nM_n/n!.\]
here, $M_n$ are the $n^{\rm th}$ moments of $f$; for this reason,
$C_f$ acts as a {\em moment generating function} for the r. v. $f$.
An important variant of this is the cumulant generating function
\[ \log C_f(\lambda)=\sum_n(i\lambda)^n\kappa_n/n!.\]
We prefer to keep the imaginary unit in these formulas, since if we
drop it the mean $C_f(\lambda)$ might not be finite.
The cumulants $\kappa_n$ are determined by induction from the system
\begin{equation}
M_n=\sum_k\sum_{{\cal I}_k}\kappa_{n_1}\ldots\kappa_{n_k}.
\label{cumulants}
\end{equation}
Here, ${\cal I}_k={\cal I}_1\cup{\cal I}_2\cup\ldots\cup{\cal I}_k$
is an arbitrary partition of $\{1,2,\ldots,n\}$ into $k$ parts,
including the identity partition, and $n_j=|{\cal I}_j|, j=1,\ldots,k$.
The condition for independence, $p_{f,g}(i,j)=p_f(i)p_g(j)$ for all $i,j$ is equivalent to
$C_{f+g}=C_fC_g$; it follows that then the cumulants of $f+g$ are the sums
of those of $f$ and $g$.

For a Gaussian distribution, all the cumulants beyond the second are zero.
There are results of the following kind: if all the cumulants
$\kappa_n$ of a distribution are zero for $n\geq N$, then they are zero
beyond $n=2$, and so the distribution is Gaussian.
These results use the positivity of the mean of a positive polynomial in $f$:
\begin{equation}
\sum_{ij}\overline{z}_iz_jM_{i+j}=E[\sum_{ij}\overline{z}_iz_jf^{i+j}]=
E[|\sum_jz_if^j|^2]\geq0.
\label{moment}
\end{equation}
Given a set of real numbers $\{M_n\}$ satisfying the positivity condition in
(\ref{moment}), it is
not obvious that $M_n$ is the $n^{\rm th}$ moment of a random variable $f$,
or that if so, $f$ is unique.
This has led to a body of work called the moment problem.

The distribution of a random variable $f$ determines that of any
differentiable function $g(f)$ of $f$; this is also a random variable;
the density of the distribution of $g$ is determined by the usual rule:
if $g$ is bijective, so that $f$ is a function of $g$,
the probability that $g$ lies between $y$ and $y+dy$
is $\rho_g(y)dy$, and this occurs if and only if $f$ lies between $x$
and $x+dx$, where $y=g(x)$. Therefore $\rho_gdy=\rho_fdx$, giving the
relation
\begin{equation}
\rho_g=|(df/dg)|\rho_f.
\label{change}
\end{equation}
If $g$ is not bijective, but has a local inverse with various branches,
$f_i$, then we have to sum over the contribution $|(df_i/dg)|\rho_{f_i}$
of each branch.

The remarkable thing is that the methods of probability theory give good
results in many cases that are not governed by chance, such as the
distribution of digits in $\pi$. Another example is the
configuration of a chaotic system at the time $t$, where $t$ is large,
given the initial configuration at time zero. If the initial
state is not specified sufficiently accurately, then the configuration at
time $t$ seems to be governed by chance, although it is not.
It was suggested by Krylov \cite{Krylov} that statistical physics is a
successful method exactly in the cases when the underlying
dynamics is chaotic. This will occur when nearby initial points
become exponentially far apart as time progresses, and this is signalled by
a positive
real part to the dominant eigenvalue of the linearised dynamics.
This largest real part is called the Lyapunov index.
We are talking here about a chaotic {\em theory};
actual experimental measurements will always have further uncertainty,
influenced by small
effects omitted from the theory. In a non-chaotic system small
forces can be omitted in the first few approximations. However, in a
chaotic system, the inclusion of one such small force can change the
outcome of the calculation at the large time $t$, making it appear to
be random. This is well modelled by omitting any attempt to include all
the actual forces, replacing those omitted by a `noise', that is, a
random term. Thus, we expect chaos to be well-modelled by a system with
increasing uncertainty, as measured by entropy. Kolmogorov, and then
Sinai took up Krylov's cause, and were able to relate the rate of
`entropy' production to the Lyapunov exponent of the dynamics.
However, Ruelle interprets this 
\cite{Ruelle} as an increase in information, available as time goes by.

Laplace's problem, of whether to assign equal probabilities to each
energy-level of a system, arises in quantum theory.
Krylov takes von
Neumann to task for assuming that the density matrix for the state of a
particle with spin produced by a quantum process should, in the absence
of any theory or experiment, be taken to be totally unpolarised. Krylov
says that this is not true for most known processes, as the polarisation
is found to be nonzero, small for some and large for others.
Krylov's view is that it should be assigned
a general density matrix; we can then estimate this matrix in the light of
experiments. This leads to the subject of {\em quantum estimation},
for which there is a body of theory. Krylov
believed that physics is not in the gambling business; we do not second
guess the state of the system and follow a strategy of hedging against
wrong guesses; rather, in physics we predict what will happen (with various
probabilities) at a later time, when the initial state is known.

Estimation theory has received an impetus from a modern development,
information theory.
Shannon introduced the entropy of the random variable $f$ taking values
$x_i$ as
\begin{equation}
S_f:=-\sum_i p_f(x_i)\log p_f(x_i).
\end{equation}
Note that
$S_f$ does not depend on the actual values that $f$ takes. The distribution
with the maximum possible entropy is easily proved to be the uniform
distribution. The school of probability known as Bayesian therefore argues
that if we know nothing whatever about $f$ it must be assigned the uniform
distribution, called the {\em prior}. Thus, Laplace's
intuition gets very respectable support. There is one big problem with this:
the uniform distribution for $f$ is not in general consistent with
the uniform distribution for say $g=f^3$, as we see from eq.~(\ref{change});
so the prior depends on the
random variable we choose to name as the one we know nothing about. This
echoes Bertrand's paradox.

A quantum version of entropy was earlier given by von Neumann.
For the classical case $(\Omega,p)$ with $\Omega$ countable it reduces to
\begin{equation}
S(p):=-\sum_\omega p(\omega)\log p(\omega).
\end{equation}
It does not make any reference to a random variable. We may obtain
Shannon's entropy of a random variable $f$ as the von Neumann entropy of
$p_f$ regarded as a probability on the space of values that $f$ takes.
Note that $S_f=S(p)$ if $f$ takes different values at
different points of $\Omega$, that is, if $f$ separates the points of
$\Omega$. We then say that $f$ is a {\em sufficient statistic}. $S_f$ is in
general less than $S(p)$, and it reduces to zero when
$f$ takes only one value. The entropies of Shannon and von Neumann are not
the same concepts, and this difference reflects their different
interpretations; the point $\omega$ is the {\em message},
and $S_f$ is the information about the message that is
on average conveyed by measuring $f$; it cannot exceed $S(p)$, which is the
entropy (missing information) in the original probability space. Naturally,
if $f$ is sure it conveys no info at all. Since 
$S_f$ depends on the random variable $f$ only through its distribution, it
has a meaning in the frequentist approach.
To compute $S(p)$, the model $(\Omega,p)$ must be given, and it
does not depend on $f$. More generally, we can define the Shannon entropy
of a set $(f_1,\ldots,f_n)$ as the von Neumann entropy of their joint
distribution on the sample space of their values. Some authors regard the
Shannon entropy as the physical entropy of a reduced
description of a physical model. The trouble with this idea
is that the introduction of noise in the measurement of $f$
causes the Shannon entropy to decrease, instead of to increase as we would
want.
 
A simple example of noise is that caused by a mapping $T:\Omega
\rightarrow\Omega$. This defines a co-action on the set of random variables:
$f\mapsto T^*f:=f\circ T$, which in fact in an endomorphism of ${\cal A}$.
If $T$ is not bijective, there might be points that can be distinguished
by measuring $f$, but not by measuring $T^*f$;
thus $S_{T^*f}\leq S_f$. This also holds more generally, when $T^*$ is a
convex linear sum of such maps, thus: $T^*=\sum\lambda_iT_i^*$. It can be
shown that this is the most general {\em stochastic map} on ${\cal A}$,
that is, linear map,
taking $I$ to $I$ and non-negative functions to non-negative functions.
The reduction in the information carried by $f$ in the presence of noise
is natural in telephony. The von Neumann entropy, on the other hand,
increases if we add noise. This is achieved by a bistochastic map $T$,
(a stochastic map whose adjoint is also stochastic). We write it as
a right action, thus: $p\mapsto pT$.
By the deep theorem
of Birkhoff \cite{Birkhoff,Ando}, a bistochastic matrix is a mixture of
permutations. Since a permutation $\Omega$ does not alter $S(p)$, and
$-p\log p$ is concave, we see that $S(pT)\geq S(p)$. Moreover,
the von Neumann entropy is not decreased by a reduced description, unlike
the Shannon version. Thus $S(p)$ is the
correct concept to represent physical entropy \cite{Streater}.

If we are given information about which $\omega$ has occurred, the
probability on $\Omega$, called the prior, changes. Suppose that $p$ is
the prior. If the information is that the sample lies in a known subset
(event) $\Omega_0\subseteq\Omega$, then Bayes's theorem on conditional
probability is used; the conditional probability is
\begin{equation}
p(E|\Omega_0):=\frac{p(E\cap\Omega_0)}{p(\Omega_0)}.
\end{equation}
This is called the posterior probability, and correctly describes the
probability among outcomes all of which lie in $\Omega_0$. A conditional
probability satisfies the axioms of probability, def.~ (\ref{probability}).

This use of information to modify the probability $p$ should not
be confused with estimation theory. There, we do not change $p$, since
after the measurement of independent samples, we continue to assume that
new samples are governed by the original $p$.
The method of estimation using the principle of maximum entropy proceeds
as follows. Suppose that we
know $\Omega$, and $f$, with $|\Omega|<\infty$ and we are also told the
average of $f$ over a number of independent trials.
We can vary $p$ over the simplex $\Sigma(\Omega)$ to find the
point that maximises the entropy of the probability, given the observed
mean value, $\eta$ say. Thus we use the method of Lagrange multipliers to
maximise
\[-\sum_\omega p(\omega)\log(p(\omega))\hspace{.4in}
\mbox{ subject to }E_p[f]=\eta.\]
Gibbs knew that the solution to this is
\begin{equation}
p(\omega)=Z^{-1}\exp(-\beta f(\omega)),
\label{Gibbs}
\end{equation}
where $Z=\sum_\omega e^{-\beta f(\omega)}$, is the Lagrange multiplier for
the normalisation condition $\sum p(\omega)=1$, and is called the
partition function. The parameter $\beta$ is the Lagrange multiplier for
the condition $p.f:=E_p[f]=\eta$, and is determined by it. Then (\ref{Gibbs})
is the least prejudiced estimate for the probability, given the mean
\cite{Ingarden,Jaynes}.

The method of maximum entropy solves an important problem in the theory of
estimation. Let $X$ be a variate of which the distribution is known
to be one of a family, ${\cal M}=\{p_\eta(i)\}_{\eta\in{\bf R}}$;
we hope to estimate
$\eta$ by measuring $X$ independently $m$ times. An {\em estimator}
$f$ is a function of the data $x_1,x_2,\ldots,x_m$
that is used for this estimate.
Thus $f$ is a function of $X$, and so is a random variable.
Since we do not know $\eta$, to be useful,
the estimator must be independent of $\eta$. We say an estimator is {\em
unbiased} if its mean is the desired parameter, thus:
\begin{equation}
p_\eta.f:=\sum_ip_\eta(i)f(x_i)=\eta.
\label{unbiased}
\end{equation}
Apart from being unbiased, a good estimator should have a small chance
of being far from the mean; so we are interested in estimators of
minimum variance, $V=p_\eta.[(f-\eta)^2]$.
To any probability $p\in{\cal M}$ define the
{\em Fisher information} as \cite{Rao,Fisher}
\begin{equation}
G=p_\eta.\left(\frac{\partial\log p_\eta}{\partial\eta}\right)^2
\label{Fisher}
\end{equation}
We recognise this as the variance of the random variable $Y=\partial\
p/\partial\eta$.
The {\em Cramer-Rao} theorem puts limits on the smallness
of the variance $V$ of an estimator $f$: 
\begin{theorem}
\begin{equation}
V\geq G^{-1}.
\label{CramerRao}
\end{equation}
\end{theorem}
For the proof, differentiate (\ref{unbiased}) with respect to $\eta$,
to get
\[\sum_i\frac{\partial p_\eta(i)}{\partial \eta}f(x_i)=1.\]
Now use $\partial/\partial\eta[\sum_ip_\eta(i)]=0,$ and rearrange, to get
\begin{equation}
\sum_ip_\eta(i)\left(\frac{\partial\log p_\eta(i)}{\partial\eta}
(f(x_i)-\eta)\right)=1.
\label{Schwarz}
\end{equation}
This is the correlation between the random variables $Y$ and $f$;
the (positive-semidefinite) covariance matrix is therefore
\begin{equation}
\left(\begin{array}{cc}
      G&1\\
      1&V
\end{array}\right)
\end{equation}
Schwarz's inequality then gives (\ref{CramerRao}).

The minimum variance allowed by (\ref{CramerRao}) occurs when the Cauchy
inequality is equality, which occurs when the factors in (\ref{Schwarz})
are proportional (with ratio dependent on $\eta$). Calling this factor
$-\partial\xi/\partial\eta$, we see that the distribution of minimum
variance must satisfy
\begin{equation}
\log p_\eta(i)=-\int^\eta \partial\xi/\partial\eta(f(x_i)-\eta)d\eta=
-\xi f(x_i)-\psi,
\end{equation}
showing that a necessary and sufficient condition is that $\{p_\eta\}$ be the
exponential family.

In the case of several parameters $\eta_1,\ldots\eta_n$, we have
estimators $f_1,\ldots,f_n$, which can
be taken to be linearly independent, but need not be functionally
independent. The state of maximum entropy, given the means $E_p[f_i]=\eta_i,
\hspace{.2in}i=1,\ldots,n$ is easily shown to be of the form
\begin{equation}
p(\omega)=Z^{-1}\exp-\{\xi_1f_1(\omega)+\ldots\xi_nf_n(\omega)\}=Z^{-1}\exp
(-f)\mbox{ say},
\label{infomanifold}
\end{equation}
where the Lagrange multipliers $\xi_i$ are determined by the given conditions
on the means. The set of probabilities of the form
eq.~(\ref{infomanifold}) form the set ${\cal M}$
called the info manifold, or the exponential family, determined by
Span$\{f_i\}$. We can regard $\{\xi_i\}$ or indeed $f$ as coordinates,
called canonical; or we can regard $\{\eta_i\}$ or indeed $p$ as coordinates,
called the expectation coordinates. In this case the Fisher {\em information
matrix} is defined to be
\begin{equation}
G^{ij}:=p.\left(\frac{\partial \log p}{\partial\eta_i}\frac{\partial\log p}{
\partial\eta_j}\right).
\end{equation}
Then the Cramer-Rao inequality (\ref{CramerRao}) becomes a matrix inequality,
where $V$ is the covariance matrix $V_{ij}:=p.[(f_i-\eta_i)(f_j-\eta_j)]$.
Equality holds only if $G=V^{-1}$, which leads to the exponential family. 

Rao showed that $G$  defines a Riemannian metric on
the tangent spaces of ${\cal M}$ \cite{Ingarden2}; as such, its components
depend on the coordinates chosen for the tangent space and it transforms
as a tensor under changes in variables.
At the point $p\in{\cal M}$, a vector in the tangent space is given in
in canonical coordinates by a random
variable $f$ in the span of the `score variables' $\hat{f}_j:=f_i-\eta_j$.
Writing $f=\sum_k\xi^k\hat{f}_k$ introduces contravariant components $\xi^k$.
These are dual to the $\eta_j$, which are covariant components.
The covariant metric is the covariance matrix
\begin{equation}
G_{ij}=G(\hat{f}_i,\hat{f}_j)=E_p[\hat{f}_i\hat{f}_j].
\end{equation}
It is the inverse of the contravariant $G^{ij}$, which explains why we get
equality in (\ref{CramerRao}).

The Massieu function $\psi:=\log Z$, where $Z$ is the partition function,
is related to the free energy; it is the generating function for
the cumulants; so we have
\begin{eqnarray}
\eta_j&=&-\frac{\partial \psi}{\partial \xi^j}\\
G_{ij}:=V_{ij}&=&\frac{\partial^2\psi}{\partial\xi^i\partial\xi^j}.
\end{eqnarray}
The entropy is the Legendre transform of $\psi$, and its second variation
is the Fisher information matrix, $G^{ij}$, the metric in the coordinates
$\eta$.

Amari showed that ${\cal M}$ is furnished with a pair of affine flat
connections, for which the global affine coordinates are $\xi^i$
and $\eta_i$ \cite{Amari}. These connections are not metric connections, but
are dual
relative to $G$. An important role in information geometry is played
by the {\em relative information} $S(p|p^\prime):=\sum_\omega p(\omega)
(\log p(\omega)-\log p^\prime(\omega))$. This distinguishes between the
points $p$ and $p^\prime$ in ${\cal M}$, in that $S(p|p^\prime)\geq 0$ and
vanishes only when $p=p^\prime$. For a modern version, see \cite{Pistone}.

The observables form the algebra ${\cal A}(\Omega)$ in which multiplication
is pointwise: $(fg)(\omega):=f(\omega)g(\omega)$; the states lie in
its dual. Thus, states and observables are not the same kind of thing,
and they transform as duals under stochastic maps.
However, states like observables are functions  of $\omega$; to distinguish
them we can write $p(\omega)$ for a state and
$(\omega)f$ for an
observable.  If $|\Omega|<\infty$, either can be identified with an element
of the formal vector space spanned by $\Omega$, thus: $\sum_\omega
\alpha(\omega)\omega\leftrightarrow\alpha$, whether $\alpha$ is regarded
as an observable or
a state. Then ${\cal M}$ is the interior of the convex hull of $\Omega$.
The permutation group of $\Omega$ acts by right action
$\omega\mapsto\omega T$. Its inverse $\omega\mapsto\omega T^{-1}$
is a co-action of the group (its product law is the opposite of that
of the group) and so can be written as a left action: $\omega\mapsto
T\omega:=\omega T^{-1}$. These induce a right action on probabilities,
and a left action on observables, by $pT(\omega):=p(T\omega)$ and
$(\omega)Tf:=(\omega T)f$, the latter written without the dual symbol $^*$.
These express associativity, as does the dual relation $pT.f=p.Tf$.

These definitions can be extended to any map $T:\Omega\rightarrow\Omega$,
whether invertible or not: we define the action on probabilities
using $T(\omega):=(\{\omega\}T^{-1})$, the inverse image of the point-set
$\{\omega\}$. Every algebraic
endomorphism of ${\cal A}$ is of the form $f\mapsto Tf$ for some map
$T:\Omega\rightarrow\Omega$, and these make up exactly
the extreme points of the convex set of stochastic maps.

In infinite dimensions, there is more than one useful topology
on the states and observables.
The modern view \cite{Pistone} is that the state $p$ and the observable
$-\log p$ are merely alternative coordinates for a point
in the info manifold. The natural class of charts are
related by monotone, convex functions, of which the stochastic maps,
\cite{Chentsov}, as well as the non-linear maps $p\mapsto-\log p$ and
$p\mapsto p^\alpha$, $0<\alpha<1$ are examples.

An active field of research is to set up quantum analogues of all this
\cite{Petz,Ray4,Ray5,Ray6} 
\section{From Bachelier to Wiener}
In 1900, Bachelier proposed a random model of the stock market
\cite{Bachelier};
the idea was that the decision to buy or sell a stock is randomly
taken by independent investors. Let us suppose that the chance $\lambda$
that the price goes up one unit $dx$ is the same as that for going down,
during any unit trading period $dt$. Let $X\in{\bf Z}$ be the random price,
and $p(x,t)$
be the probability that the price is $x$ at time $t$; then the
new probability $p(x,t+dt)$ can be unchanged, or can change due to
a movement down from $x+dx$ or a movement up from $x-dx$. The probabilities
of these are, respectively, $1-2\lambda$, $\lambda$ and $\lambda$. Thus we
get the relation
\begin{equation}
p(x,t+dt)=(1-2\lambda)p(x,t)+\lambda p(x+dx,t)+\lambda p(x-dx,t).
\label{Bachelier}
\end{equation}
Let $T$ be the tridiagonal infinite matrix $\{\lambda,1-2\lambda,\lambda\}$.
Then the row $T_{xy},y\in{\bf Z}$ is the conditional probability, that the
price will be $y$ at time $(N+1)dt$, given that it is $x$ at time $Ndt$.
In fact, $T$ is a stochastic matrix, which happens to be symmetric.

Suppose that at $t=0$ the price is $x_0$; then in time $Ndt$, the
price will follow the path $\gamma:=x_0\mapsto x_1\mapsto\ldots\mapsto x_N$
with probability 
\begin{equation}
p(\gamma)=T(x_0,x_1)T(x_1,x_2)\ldots T(x_{N-1},x_N).
\label{path}
\end{equation}
This is
called the random walk on ${\bf Z}$ determined by $T$, starting at $x_0$.
The set of allowed paths starting at $x_0$ is a finite subset of
$\Omega={\bf Z}^N$. $p(\gamma)$ is a probability on $\Omega$, and
the structure is called a Markov chain. An alternative point of view is to
start with $p_0\in\Sigma({\bf Z})$,
and to follow the path in $\Sigma({\bf Z})$ given by the time evolution.
By Bayes's law, the probability that at time $t=1$ the particle is at $x_1$
whatever its initial position, is
$p(x_1,1)=\sum_{x_0}p(x_0,0)T(x_0,x_1)$; this can be written as the matrix
product $p_0T$, where $p_t$ is a row vector made from the components of
$p(x_t,t)$.
By induction, the probability that at time $N$ the particle is at $x$
is $p_0T^N$. In this way, a Markov chain is described by a semi-group
of stochastic maps $T(t):=T^t$ acting on $\Sigma(\Lambda)$. Obviously
\begin{eqnarray}
T(0)&=&1\\
T(s)T(t)&=&T(s+t),\hspace{.4in} s,\;t\in {\bf N}.
\end{eqnarray}
One of the themes of probability theory is the relationship between a
semi-group of stochastic maps and a probabiliy on the corresponding
path space. The latter is called a {\em dilation} of the former.
Since $T$ is independent of time, the chain is said to be {\em
stationary}; if we limit the allowed space to be a finite set $\Lambda
\subseteq{\bf Z}$, we get a finite Markov chain, in which case
there is at least one stationary distribution $p^*$; this means that
$p^*T=p^*$, so that 1 is a left eigenvalue of $T$. If some power of $T$
has all its matrix elements positive, then the Perron-Frobenius theorem
tells us that $1$ is a simple eigen-value, and all the others have
modulus less than $1$. One can then show that $pT^n\rightarrow p^*$
as $n\rightarrow\infty$; the system converges exponentially
to equilibrium. We then say that the dynamics is mixing.
There are similar results in infinite dimensions, but
to get exponential convergence we need to show that there is a {\em
spectral gap}. This means that $1$ is simple and lies a finite distance
from the next eigenvalue of $TT^*$. To prove this in the case at hand is
usually the key to the study of the long-time behaviour. 
The Markov property is
that the probability of getting to $x$ at time $t+1$ depends only on
where the particle was at time $t$, and not on the previous path.
The study of Markov chains was started in the $19^{\rm th}$ century,
and is a huge subject.

Fick obtained an equation similar to (\ref{Bachelier}) for the
diffusion of particles in one dimension.
If $dx$ and $dt$ become small such that $(dx)^2/dt\rightarrow a$, a finite
limit, we say the system is following the diffusion limit. Rearranging,
and taking the diffusion limit,
Fick obtained the heat equation for the probability density, which we call
$\rho$:
\begin{equation}
\frac{\partial \rho}{\partial t}=\kappa\frac{\partial^2\rho}{\partial x^2}.
\label{heat}
\end{equation}
Here, $\kappa=a\lambda$. This is not a very good model of the market; apart
from the omission of drift, the gains in price should grow with the
overall price. As it is, negative prices are possible.

The heat equation (\ref{heat}) can be written
in the form of a conservation law:
\begin{equation}
\frac{\partial \rho}{\partial x}+\mbox{div}\,j(x,t)=0
\end{equation}
where $j(x,t)=-\kappa\nabla \rho$. At this stage,
mathematicians did not have the continuous version of the sample space
$\Omega$; this was to be Wiener's great construction.

In his celebrated work of 1905 \cite{Einstein1}, Einstein also used
(\ref{heat}) to describe the Brownian motion of small particles in a warm
liquid. He was mindful of Stokes's law of diffusion; this says that in a
viscous liquid a small particle under a constant force, such as gravity,
will increase its speed towards a terminal velocity $v$ say, which is
proportional to the force. Einstein required that in equilibrium
the current $v\rho$ due to this flow should balance the diffusion due to the
density gradient, so that steady state should obey
\begin{equation}
-\kappa\nabla \rho+v\rho=0.
\end{equation}
The solution to this in the case of gravity, where $v=-|v|$ in the
$z$-direction, is
\begin{equation}
\rho(x,y,z)=\mbox{const.}e^{-|v|z/\kappa},
\end{equation}
and this should be the 
Maxwell-Boltzmann law at the temperature $\Theta$ of the liquid,
\begin{equation}
\rho(x)=Z^{-1}e^{-mgz/(k_B\Theta)}.
\end{equation}
Einstein thus obtained the famous Einstein relation
\begin{equation}
F=k_B\Theta v/\kappa.
\end{equation}
His treatment is not
complete, since he omitted
the drift term in the diffusion equation! See \S 4, (1) in \cite{Einstein1}.
In a detailed
study, Smoluchowski \cite{Smol} wrote down the diffusion equation with drift
\begin{equation}
\frac{\partial \rho}{\partial t}=\kappa\frac{\partial \rho}{\partial x^2}-
v\frac{\partial \rho}{\partial x}.
\label{heat2}
\end{equation}
now known as the Smoluchowski equation, and is a special case of the
Fokker-Planck or backward Kolmogorov equation. He solved this by
using the method of images for several systems with
boundaries, such as the mass of air above the ground, and obtained
the approach to the stationary state expected by Einstein. 

It was known that one can solve eq.~(\ref{heat2}) exactly,
to fit a more or less arbitrary initial function $\rho(x,0)=f(x)$, by using
the Green function (in one dimension)
\begin{equation}
G(x,t):=[4\pi\kappa t]^{-(1/2)}e^{-(x-vt)^2/(4\kappa t)}.
\end{equation}
This satisfies eq.~(\ref{heat2}), and converges in the sense of
distributions to the Dirac $\delta$-function as $t\rightarrow0$. Then
\begin{equation}
\rho(x,t)=\int_{-\infty}^\infty G(x-y,t)f(y)\,dy
\label{solution}
\end{equation}
satisfies eq.~(\ref{heat2}) and the boundary condition.
The operator whose kernel is $G$ is the continuum analogue of the matrix
$T^n$ of the Markov chain, 

When the force and
temperature are slowly varying, we get the coupled system
\begin{eqnarray}
\frac{\partial \rho}{\partial t}&+&\mbox{div}\,J=0;\\
\frac{\partial\Theta}{\partial t}&=&\kappa^\prime \mbox{div}\,\nabla\Theta+\kappa
{\bf F}.{\bf F}/k_B\Theta.
\end{eqnarray}
Here, $J(x,t)=-\kappa\nabla(\rho+V\rho/\Theta)$, where $V$ is the potential
giving rise to the force $F$. The source term in the heat equation is $F.J$,
the power of the external force supplied to the particle, all of which is
converted into heat. This system obeys
the first and second laws of thermodynamics \cite{RFS1}.

Consider now the solution (\ref{solution}) to (\ref{heat2}). Because $G$ is positive,
the density remains positive for all time, and the conservation law shows
that the integral of $\rho$ over space is constant. So we get a flow
through the space of probabilities. The question arises, is there a
process in continuous time associated with the Smoluchowski equation?
The answer is yes, and this was the result of the work of Wiener,
and later, Ito. An alternative idea was introduced by Langevin, who
considered Newton's laws, in which a part of the external
force, denoted $F$, is random; friction enters as a damping
force proportional to the velocity, parametrised by $\gamma>0$.
Thus his equation is
\begin{equation}
\frac{d^2x}{dt^2}=-\frac{\partial V}{\partial x}-\gamma\dot{x}+F(t).
\label{Langevin}
\end{equation}
This is the equation for a single particle, but as $F$ is random, the
position
$x(t)$ becomes random as time goes by, even if its initial condition is
given. Statistical properties of $x$ are determined by those of $F$; the
relation of these to the Smoluchowski equation were studied by Fokker and
Planck, but were fully understood only in terms of stochastic calculus.
One might assume that $F$ is Gaussian distributed, and is of mean zero,
with independent values at different times. This would now be described as
{\em white noise}. Langevin's work started the enormous field of
stochastic differential equations.

In 1904 Lebesgue tried to set up a general theory in which every subset
of $[0,1]$ is assigned a measure.
\cite{Legesgue}. 
The very next year, G. Vitale showed that the scheme was inconsistent
\cite{Vitale}. Hausdorff \cite{Hausdorff} and Banach and Tarski,
showed that the measure could not be additive \cite{Williams}. The point
is that some
sets are so bad they cannot be assigned a measure, even a finitely
additive one. This led to the concept of measurable set. Let us start with
the Borel measurable sets on $[0,1]$.

Let $\Omega=[0,1]$; let us say that a collection ${\cal B}$ of subsets
of $\Omega$ form a {\em tribe} if
\begin{enumerate}
\item $\Omega\in{\cal B}$
\item whenever $B\in{\cal B}$, we have $B^c:=\Omega-B\in{\cal B}$;
\item whenever $A\in{\cal B}$ and $B\in{\cal B}$, we have $A\cup B\in
{\cal B}$.
\end{enumerate}
Such a collection of subsets is also called a Boolean ring, or a Boolean
algebra. The collection ${\cal B}$ is
actually a ring, with multiplication given by intersection, and addition
given by symmetric difference, that is $A+B:=A\cup B-A\cap B$. It is
also an algebra in the technical modern sense, but trivially in that any
ring is an algebra over the field
consisting of two numbers, $0$ and $1$. Since this ring structure plays no
role in the theory, we prefer not to furnish ${\cal B}$ with the extra
structure `+', and will use the word `tribe' instead.

We define a
$\sigma$-tribe to be a collection ${\cal B}$ of sets $B_i\subseteq\Omega$
such that 3. above is replaced by\\
$\;\;3_\infty.\mbox{ if }B_i\in{\cal B} \mbox{ is a countable family of disjoint
sets, then }\cup B_i\in{\cal B}.$\\
The set of all subsets of a set $\Omega=[0,1]$ is obviously a $\sigma$-tribe,
and indeed satisfies uncountable additivity as well. This $\sigma$-tribe is
called the {\em power set} of $\Omega$. But, as we saw, there are no
useful definitions of measure on the power set. Another easy case is the
collection of all countable subsets of $\Omega$: the union of
a countable collection of countable subsets is countable. However, any
countable set has length zero, since it can be covered by a sequence
of intervals of length $\leq\epsilon/2,\;\epsilon/4\;\epsilon/8,\ldots$, of total
length $\epsilon$. Since $\epsilon$ can be anything, the set has length zero.
To get some sets of non-zero length, let us consider the tribe ${\cal B}_0$
of all finite disjoint unions of open, closed and half-open intervals. We
could add to ${\cal B}_0$ all countable unions of sets in ${\cal B}_0$, and
all complements in $\Omega$ of sets in the tribe so obtained. Call this
${\cal B}_1$. Then we would need to consider the
collection of countable unions of sets in ${\cal B}_1$, and their
complements, to get a new tribe ${\cal B}_2$, and so on. Does this end
up with a well-defined $\sigma$-tribe?
The following argument does the trick. Let ${\cal G}$ be any $\sigma$-tribe
containing all sets in ${\cal B}_0$, and let $C$ be the set of all
such $\sigma$-tribes. Then $C$ is non-empty, as it
contains the power set at least. Then form
\begin{equation}
{\cal B}=\bigcap_{{\cal G}\in C}{\cal G}.
\end{equation}
That is, ${\cal B}$ contains those subsets of $\Omega$ that lie in all
$\sigma$-tribes ${\cal G}$, and no other subsets. In particular, ${\cal B}$
contains all subsets in ${\cal B}_0$, ${\cal B}_1$ etc. In fact, by using
the techniques of set theory, one can prove that ${\cal B}$ is
smallest $\sigma$-tribe containing all the open
intervals in $\Omega=[0,1]$; it is called the {\em Borel tribe}. One can ask
whether we have arrived at the power set after all, or have something
without the pathological sets. That ${\cal B}$ contains only nice sets
follows the construction of a countable measure on its sets, namely
the Lebesgue measure.

A {\em finitely additive measure} on a tribe ${\cal B}$
is a map $\mu:{\cal B}\rightarrow {\bf R}_+\cup\{+\infty\}$ such that
\[\mu(A\cup B)=\mu(A)+\mu(B)\mbox{ for all disjoint }A, B\in{\cal B}.\]
If $\mu(\Omega)=1$, it is a finitely additive probability measure.
To do analysis, we must be able to take some limits, and so we
now assume that ${\cal B}$ is a $\sigma$-tribe.

A probability measure on $(\Omega,{\cal B})$ is a map $\mu:{\cal B}
\rightarrow{\bf R}^+$ such that
\begin{enumerate}
\item $\mu(B)\geq0$ for all $B\in{\cal B}$;
\item $\mu(\Omega)=1$;
\item if $B_i$ is a countable collection of disjoint sets in ${\cal B}$,
then
\[\mu(\cup B_i)=\sum_i\mu(B_i).\]
\end{enumerate}
Considering the tribe ${\cal B}_0$ of finite unions of disjoint open,
closed
and half-open intervals, we can define the {\em Lebesgue measure} of
$B\in{\cal B}_0$ to be the sum of the usual lengths of the intervals
involved. It is then proved that there is a countably additive measure
on the Borel $\sigma$-tribe, which agrees with the length on the intervals.
This measure is called the {\em Lebesgue} measure.

It is sometimes useful to extend the concept of measure to unbounded sets
such as ${\bf R}$, whose total length is infinite. For this, we just
drop axiom 2. above.

So much for the measure; integration theory needs a remark as well.
Suppose that we have a function $y=f(x)$, where $x\in[0,1]$ and
$y$ is real-valued and bounded,
and we seek a way of finding the area under the graph of $y$ against
$x$. In Riemann's {\em method of integration} we divide the $x$-axis
into a large number of small intervals, $[0,x_1],\,(x_1,x_2],\ldots,
(x_N,1]$, and define $y_i$ to be the smallest value of $y$ in the
interval $(x_i,x_{i+1}]$ and $Y_i$ to be the largest value.
Now define the two approximations
to the area, known as the upper sum and the lower sum,
$R^+=\sum_i Y_i(x_{i+1}-x_i)$ and $R^-=\sum_i y_i(x_{i+1}-x_i)$.
As we refine the subdivision, $R^+$ decreases
and $R^-$ increases. If the limits of these are equal, we say that the
function is Riemann-integrable, and take
their common value as the area under the curve $y=f(x),\;\;0\leq x\leq1$.
One shows that continuous functions are
integrable, and can establish the fundamental theorem of the
calculus; a generalisation, called the Riemann-Stieltjes integral,
can be defined, if we replace $x_{i+1}-x_i$ by $P(x_{i+1})-P(x_i)$, where
$P$ is an increasing function of bounded variation, continuous from the left.
We write the integral as $\int y(x)dP(x)$.
To define the integral of unbounded functions, various limiting methods were
invented. The theory is not really satisfactory.

Lebesgue introduced a new form of integration: compared
with Riemann's method, it is done
the other way round. As the first step, only positive functions are
considered. Then, we divide the $y$-axis into intervals $([0,y_1],
\,(y_1,y_2],\ldots,(y_N,\infty))$, and for each interval, look for the
inverse image of each interval under the map $f$. That is, we
consider the subset of the $x$-axis consisting of $x$ such that $f(x)\in
(y_i,y_{i+1}]$. This set, denoted by $f^{-1}(y_i,y_{i+1}]:=B_i$, may
consist of many pieces, and so will not
always be an interval. We require, however that it should be a set in
the Borel $\sigma$-tribe ${\cal B}$; if this holds for every
subdivision of the $y$-axis into intervals, we say that the function $f$
is ${\cal B}$-{\em measurable}. The set $B_i$ will have a `length',
namely, its Lebesgue measure, $\mu(B_i)$.
We approximate the area under the graph of $f$ 
by the sum
\[ L(f):=\sum_i y_i\mu(B_i).\]
This is positive and increases as we refine the partition
of the $y$-axis. If its supremum over all partitions is finite, we say
that $f$ is Lebesgue-integrable, and write
\begin{equation}
\int_0^1 f(x)dx=\sup L(f).
\end{equation}
We can integrate functions that are not positive, provided that the
positive and negative parts are separately integrable, and we integrate
complex functions by treating the real and imaginary parts separately.
This generalises the Riemann integral in that any Riemann-integrable
function is Lebesgue integrable, and then both versions give the same
answer. 

Lebesgue integration has the following easy generalisation, which is
important for
probability. Suppose that $\Omega$ is any set, provided with a
$\sigma$-tribe ${\cal B}$; the pair $(\Omega,{\cal B})$ is called
a {\em measurable space}.
A real-valued function is said to be ${\cal B}$-measurable
if the inverse image of every open interval lies in ${\cal B}$:
\[f^{-1}(y_1,y_2):=\{\omega\in\Omega:y_1<f(\omega)<y_2\}\in{\cal B}.\]

A random variable is then simply a real-valued ${\cal B}$-measurable
function on $\Omega$. Given a measure $\mu$ on $(\Omega,{\cal
B})$, not necessarily of finite total measure, we can
regard as the same random variable $f$ two that differ only on
a set of $\mu$-measure zero; they are called {\em versions} of $f$.
The set of all bounded random variables forms a commutative {\em
algebra} ${\cal A}(\Omega)$ with norm $\|f\|_\infty:=\inf\,\sup_\omega
|f(\omega)|$; here, $\inf$ is taken over all versions of $f$.
The sets in ${\cal B}$ are called events. The integral
of a positive measurable function (with respect to the measure $\mu$) is
defined
similarly to the case when $\Omega={\bf R}$. If $\mu$ is a probability
measure, this integral is called the mean $\mu.f$ of $f$ in the state $\mu$.
and if $\mu.|f|<\infty$ we write $f\in L^1(\Omega,{\cal B},\mu)$.
More generally, we write $f\in L^p(\Omega,{\cal B},\mu)$, $1\leq p<\infty$
if $f$ is ${\cal B}$-measurable and $|f|^p$ is integrable. These are Banach
spaces with norm $\|f\|_p:=(\int |f(\omega)|^p\,d\mu)^{1/p}$.
The probability of an event $B$ is taken to be $\mu(B)$.
Each measure $\mu$ defines an element
of the dual space of ${\cal A}$ by the linear form $f\mapsto \int f\,d\mu$.

We have remarked that the original motivation for introducing the
$\sigma$-tribe was to avoid pathology. However, the concept has been very
useful in a heuristic way, to describe the {\em information}
carried by events and observables
in a random theory based on a measure space $(\Omega,{\cal B},\mu)$;
in particular, it is useful to consider a sub-tribe
or sub-$\sigma$-tribe, of ${\cal B}$. Suppose that $B\in{\cal B}$ is an
event; it is determined by its {\em indicator function}, $\chi_{_B}(\omega)$
which is $1$ if $\omega\in B$ and zero outside $B$. If $\mu(B)\neq 0,1$
and $f$ is measurable, we can define the conditional expectation
\begin{equation}
E[f|B]=\sum_\omega f(\omega)\mu(\omega|B).
\end{equation}
We may also find the conditional probability of $A\in{\cal B}$, given
that $B$ did not happen: $\mu(A|B^c)=\mu(A\cap B^c)/\mu(B^c)$, and the
corresponding conditional expectation
\begin{equation}
E[f|B^c]=\sum_\omega f(\omega)\mu(\omega|B^c).
\end{equation}
We may regard the pair of numbers, $\{E[f|B],E[f|B^c]\}$ as defining
a simple measurable function on $\Omega$, equal to $E[f|B]$ if
$\omega\in B$ and to $E[f|B^c]$ if $\omega\notin B$. Let us now
generalise this idea. Let $B_1,\ldots,B_n\in{\cal B}$ be
disjoint measurable sets such that $\mu(B_j)\neq 0$ for all $j$, and
$\mu(\cup_j B_j)=1$. These sets generate a tribe, say ${\cal B}_0$
(by various unions; there are $2^n$ such unions). If $f$ is measurable,
the functions on $\Omega$ defined by
\begin{equation}
F_f(\omega)=E[f|B_j]\;\mbox{ if }\omega\in B_j
\end{equation}
are measurable relative to ${\cal B}_0$. They take constant values,
$E[f|B_j]$ on each $B_j$ and so can be written
\begin{equation}
F_f(\omega)=\sum_j\chi_j(\omega)c_j\;\mbox{ where }c_j=E[f|B_j].
\end{equation}
Conversely, every function $F$, measurable relative to ${\cal B}_0$,
has this form for some $\{c_j\}$. The map, $f\mapsto F_f$, is linear
and is called the {\em conditional expectation} of $f$ given
${\cal B}_0$. This map leaves invariant the
vector space of ${\cal B}_0$-measurable functions, and indeed
is the orthogonal projection of $L^2(\Omega,{\cal B},\mu)$ onto
$L^2(\Omega,{\cal B}_0,\mu)$.

The tribe ${\cal B}_0$ tells how
fine was the division into the sets $B_j$, and determines how
much detail can be obtained from the functions that are
${\cal B}_0$-measurable. From the fact the $F$ is the orthogonal projection,
we see that $E[f|{\cal B}_0]$ is the best approximation (in the $L^2$-sense)
to $f$ by functions that are ${\cal B}_0$-measurable.

Consider for example the price
of a stock at time $t$, where $t$ is a non-negative
integer; $S(t)_{t\geq0}$ are then a family of
random variables on $\Omega$, and while we can find out the prices
up to the present time, we cannot know the future. Suppose that $t=N$ is
the present. The {\em information}
contained in the knowledge of the prices at $N+1$ previous times, namely
$S(t=0)=s_0,\,S(t=1)=s_1,\,\ldots,S(t=N)=s_N$, selects in $\Omega$ a
particular level set of these functions: this is the event
\[\{\omega\in\Omega:S(0)(\omega)=s_0\ldots S(N)(\omega)=s_N\}\]
Since we assume that $S(t)$ are ${\cal B}$-measurable, this set lies in
${\cal B}$, by the intersection property. The same for any other possible
set of values of these observations. There is a smallest $\sigma$-tribe
with respect to which all these functions are measurable, and in fact,
this $\sigma$-tribe is generated by all the level sets described above.
Call this ${\cal B}_{\leq N}$. The set of all random variables that
are ${\cal B}_{\leq N}$-measurable is exactly the set of functions of the
data $S(0),\ldots,S(N)$, measurable in the Lebesgue sense; they can
therefore be computed from the data we have access to.

The increasing family $\{{\cal B}_{\leq n}\}$ is called the filtration
generated by the process.
It provides a neat formulation of the Markov condition for a process $X_n$;
let ${\cal B}_n$ be the $\sigma$-tribe generated by the r. v. $X_n$.
Then a process is called Markovian if
\begin{equation}
E[X_n|{\cal B}_{\leq m}]=E[X_n|{\cal B}_m]\mbox{ if }n\geq m.
\end{equation}
The idea is that the information contained in $X_m$, the present value,
tells us as much about the future as the whole previous history.
Consider again the semigroup $\{T^n\}$ of stochastic maps, acting on
$\Sigma({\bf Z})$ in one time-step as in eq.~(\ref{Bachelier}).
One can check that if $p_0$ is the initial probability distribution of
the initial point of the path, then
\begin{equation}
p_0T^n=E_{p_n}[x_n|{\cal B}_0].
\label{KBE}
\end{equation}
Here, $\gamma=(x_0,\ldots,x_n)$ and $p_n(\gamma)=p_0(x_0)p(\gamma)$
where $p(\gamma)$ is as in (\ref{path}).

Wiener \cite{Wiener} was able to put the Bachelier-Einstein diffusion
theory on a rigorous footing. He has to define, first, the sample space
$\Omega$; then
he needs a $\sigma$-tribe ${\cal B}$ and a measure on it; he also
needs a family of ${\cal B}$-measurable functions $X_t(\omega)$ whose
distribution has density of probability equal to $\rho(x,t)$ obeying the
diffusion equation. Finally, he needs to get the continuum version of
eq.~(\ref{KBE}).

Let $\Omega$ be the set of all continuous functions $\omega$ of $t\geq 0$
with $\omega(0)=0$;
these are called `Brownian paths'. Let $(x_1,y_1)$ be an interval of the
real line, which we call a {\em gate}; we now
consider the subset of paths which pass through the gate at time $t_1$. This set is called
the {\em cylinder set based on} $(x_1,y_1)$.
In symbols, it is
\[\{\omega\in\Omega:x_1<\omega(t_1)<y_1\}\]
The $\omega(t)$ for various $t$ are
coordinates of the point $\omega$; we have a condition
on only one of the coordinates; the rest run over the real line. Consider
another cylinder set, similarly constructed at time $t_2>t_1$, based on
another open interval $(x_2,y_2)$. The intersection of these sets is a
cylinder set based on rectangle $(x_1,y_1)\times(x_2,y_2)$ in the plane
made by the coordinates $\omega(t_1),\omega(t_2)$. The path $\omega(t)$
passes through the first gate at time $t_1$ and the
second at time $t_2$; it is a slalom. Consider the collection of subsets
of $\Omega$ consisting of all these cylinder sets defined by slaloms with
any finite number of gates, at any selection of different positive times.
The finite unions of these form a tribe. The smallest $\sigma$-tribe
${\cal B}$ containing all these is the one we choose, so obtaining
the measurable space $(\Omega,{\cal B})$.

We first define a finitely additive measure on the tribe of cylinder sets.
It is enough
to give the measure of a general cylinder set, and to use the finite
additivity. Starting at $x=0$, the probability density
that a diffusing particle reaches $x_1$ at time $t_1$ is taken to be the
Gaussian given by the Green function; thus the probability of lying in the
interval $x_1,y_1$ is
\begin{eqnarray}
\mbox{Prob}\{\omega(t_1)\in(x_1,y_1)|\omega(0)=0\}&=&\frac{1}{(4\pi\kappa t_1)
^{1/2}}\int_{x_1}^{y_1}e^{-x^2/(4\kappa t_1)}dx\nonumber\\
&=&\int_{x_1}^{y_1}G(x,t_1)dx.
\label{gate1}
\end{eqnarray}
The probability that the path goes through two gates, $(x_1,y_1)$ at $t_1$
and $(x_2,y_2)$ at $t_2$ is defined to be
\begin{eqnarray}
\mbox{Prob}\{\omega(t_1)&\in&(x_1,y_1)\mbox{ and }\omega(t_2)\in(x_2,y_2)|
\omega(0)=0\}\nonumber\\
&=&\int_{x_1}^{y_1}dx\int_{x_2}^{y_2}dx^\prime G(x,t_1)G(x^
\prime-x,t_2-t_1).
\label{gate2}
\end{eqnarray}
This can be interpreted as Bayes's theorem, in which $G$
is the conditional probability density.
Similarly, the probability of any cylinder set, based on a finite set of
gates, can be given. The probability is the same, whether the gates are
open, closed or half-open. We would like the measure we are constructing
to be at least finitely additive. Thus we take the measure of the union
of two disjoint cylinder sets to be the sum of the measures we have
just given them individually. A possible problem arises if we add together
infinitely many gates at time $t_1$ to make up the whole line; for, we
would like our measure to be countably additive, and we need the {\em
consistency} condition between the two ways to define the probability
of reaching the gate $(x_2,y_2)$: from $0$ directly, with no gate at $t_1$,
as given by eq.~(\ref{gate1}), 
or as the sum over all paths going through any complete set of disjoint
gates at $t_1$, as got by summing eq.~(\ref{gate2}). Indeed, we do
get the same answer, because of the propagating property of $G$:
\begin{equation}
\int_{-\infty}^{\infty}dx^\prime\;G(x-x^\prime,t_1)G(x^\prime-y,t_2-t_1)=
G(x-y,t_2).
\end{equation}
This is a continuous version of the obvious property of the stochastic
matrices $T^n$ of a Markov chain, namely $T^mT^n=T^{m+n}$, in which the
matrix product, expressed as the sum over an intermediate index, is
replaced by the integral over the point $x^\prime$. Thus our equation
just expresses the semi-group property of the time-evolution of a
first-order equation, here the heat equation.
It is seen here as the main point which establishes the additivity
of the finitely additive measure we have constructed on the tribe of
cylinder sets.

Let us define ${\cal B}_{[s,t]}$ as the $\sigma$-tribe generated by the
cylinder sets labelled by times in the interval $[s,t]$, and ${\cal B}$
that generated by all of these. Then Wiener proved that there exists
a unique measure on the measurable space $(\Omega,{\cal B})$ that coincides
with the measure above on the tribe of unions of such cylinder sets.
This measure is now called {\em Wiener measure}.
The Wiener process starting at $0$ is then the family of random variables
defines by $W_t(\omega):=\omega(t),\hspace{.1in}t\geq0$.
The process has the following properties:
\begin{enumerate}
\item $W_t-W_s$ is Gaussian with mean zero and variance $t-s$, for $t>s$.
\item $W_t-W_s$ is independent of $W_v-W_u$ if $0\leq u\leq v\leq s\leq t$.
\item $W_0=0$.
\end{enumerate}
These properties characterise the process.
By requiring that $W^x_0=x$ we get the Wiener process $W_t^x$ 
starting at $x\in{\bf R}$.

We now need the concept of the {\em symmetric Fock space} $\Gamma({\cal H})$
of a Hilbert space ${\cal H}$. The $n$-fold tensor product $\otimes^n{\cal
 H}$ is the completed span of the symbols $\otimes_{i=1}^n\psi_i=\psi_1
\otimes\ldots\psi_n$ with the scalar product
\[ \langle\otimes\psi_i,\otimes\phi_i\rangle:=\prod_{i=1}^n
\langle\psi_i,\phi_i\rangle.\]
The symmetric tensor product ${\cal H}^n:=\otimes_S^n{\cal H}$
is the subset of symmetric
tensors, called the $n$-particle space; the zeroth tensor power is taken to
be ${\bf C}$. The Fock space
$\Gamma({\cal H})$ is the direct sum $\oplus_{n=0}^\infty{\cal H}^n$.
This has the functorial property
\[ \Gamma({\cal H}_1)\otimes\Gamma({\cal H}_2)=\Gamma({\cal H}_1)
\oplus\Gamma({\cal H}_2).\]
As a special case, $\Gamma({\bf C})={\bf C}\oplus{\bf C}\oplus\ldots$.

There is a unitary map $L^2(\Omega,{\cal B},\mu)\rightarrow\Gamma
(L^2([0,\infty),dt)$, in such a way that $\oplus_{j=0}^n{\cal H}^j$
is identified with the $L^2$-completed span of the polynomials in $W_t$
of degree $\leq n$ \cite{Segal}. The $n$-particle space is then identified successively
by Gram-Schmidt orthogonalisation with the part orthogonal to the
$k$-particle spaces, $k<n$. This is Wiener's {\em chaos expansion}
\cite{Wiener2}.
In particular, the one-particle space is spanned by $\{W_t\}_{t\geq0}$.

For each fixed $t$, the space $L^2(\Omega,{\cal B},\mu)$ contains the
random variables $I$, $W_t$, ...$W^n_t$...They can act as multiplication
operators successively on the vector $1$, to get $n$ vectors.
Suppose we orthogonalise them by
the Gram-Schmidt procedure. Since $W_t$ is Gaussian, we get the Hermite
polynomials in successive spaces, and any $L^2$ function of $W_t$
has a convergent expansion as a sum of its components in these spaces.
The subspace we get can
be identified as the Fock space over the one-dimensional space spanned by
$W_t$. We shall see that these  polynomials are Wick-ordered powers
\cite{Wick} of $W_t$, and that they are martingales.

Now suppose that $u>s$; since $W_u$ is independent of $W_u-W_s$, and they
are Gaussian, they are orthogonal in the one-particle space, which
can thus be written as the direct sum $L^2([0,\infty),dt)=L^2([0,s),dt)
\oplus L^2([s,\infty),dt)$. By the functorial property of Fock space, we therefore
can write 
\begin{equation}
L^2(\Omega,{\cal B},\mu)=L^2(\Omega,{\cal B}_{[0,s]},\mu)\otimes
L^2(\Omega,{\cal B}_{\geq s},\mu).
\end{equation}
We can similarly split Fock space into arbitrarily many factors,
corresponding to any partition of the time axis into intervals:
it has the property of a continuous tensor product.

The continuous analogue of the semigroup $(\gamma T^n)_m:= \gamma_{m+n}
,\;m,n=0,1,2\ldots$ of the random walk is the {\em left-shift} of the paths:
$(\omega T_s)(t)=\omega(s+t)$. This induces the dual action on the
observables:
\begin{equation}
T^*_s:L^2(\Omega,{\cal B},\mu)
\rightarrow L^2(\Omega,{\cal B}_{\geq s},\mu),\;s\geq 0.
\label{shift}
\end{equation}
\noindent This operator is isometric but not invertible. We can also embed
$L^2(\Omega,{\cal B},\mu)$ in the two-sided space
$\Gamma(L^2(-\infty,\infty))$, on which the left shift is unitary, and
induces the action of the group ${\bf R}$ rather than the semigroup
${\bf R}^+$.
In that case the paths are not conditioned to pass through the origin, and
only the differences, $W_t-W_s$ make sense as vectors or operators.

\section{The Quantum Leap}
The remarkable discovery of matrix mechanics by Heisenberg in 1925
is comparable to that of the theory of relativity in 1917.
Clifford had
speculated that the world might have chosen a geometry other than Euclidean.
It was agreed that it was an experimental question, and that the data
agreed with Einstein's theory. Though the classical axioms were yet to
be written down by Kolmogorov, Heisenberg, with
help of the Copenhagen interpretation, invented a generalisation of the
concept of probability, and physicists showed that this was the model of
probability chosen by atoms and molecules.

According to Einstein et al. \cite{Einstein2} a concept
is deemed to be an {\em element of reality} within
a specified theory if there is a mathematical object in the theory which
is assigned to the concept, and which takes a definite value (when the
state of the system is given). This
is now called an {\em observable}.
For example, the choice of the zero-level of a potential function, is not
an observable since it is not determined by the state of the system.
They are not here discussing
random samples, which at the time would have been described as an ensemble.
In that case, they might have conceded that a concept could be regarded as
an element of reality if, in a random selection of the system from an
ensemble, there is a definite random variable assigned to the physical
concept.
The interpretation of a theory is not complete unless it is
specified at the outset which mathematical
objects arising in the theory correspond to observables. Thus in a theory
with randomness in classical physics, there is a space $(\Omega,{\cal B})$
and an observable is a random variable, and an ensemble is a probability
measure on $\Omega$. A non-random state is given by a point-measure. In
this state any r. v. has zero variance, thus satisfying {\em EPR}.

In quantum mechanics, this is not the case; an
observable is a Hermitian matrix $A$, or in modern terms, a self-adjoint
operator on a given Hilbert space ${\cal H}$; the possible values one can
find in a measurement are the eigenvalues of $A$.
A wave-function is determined by a vector $\psi\in{\cal H}$; but only
unit vectors are used, and $e^{i\theta}\psi$ represents the same state as
$\psi$. Thus the state is the equivalence class
$\{\psi\}=\{e^{i\alpha}\psi,\alpha\in{\bf R}\}$. If
$\dim{\cal H}=n<\infty$, such equivalence classes make up the
projective space $CP^{n-1}$. An element of $CP^{n-1}$ determines the
expectation value of any observable $A$ by $\langle\psi,A\psi\rangle$,
which according to the Copenhagen interpretation, is the mean value of $A$
if measured many times in the state $\{\psi\}$. It is seen to be
independent of the representative vector $\psi\in\{\psi\}$. Such a state is
called a {\em vector state}.
The concept of state was generalised by von Neumann to
include random mixtures of vector states . Let ${\cal B}({\cal H})$
denote the set of bounded operators on ${\cal H}$; this is a complex vector
space, and also $^*$-algebra, where conjugation is given by the adjoint
and multiplication is the usual product of operators.
A state is given by a positive operator
$\rho$ of trace 1, called a density operator,
and the expectation of an observable $A$
is taken to be $m_1(A):=\mbox{Tr}\,(\rho A)$. Any density operator
determines an element of the dual space to ${\cal B}({\cal H})$
by the map $A\mapsto m_1(A)$.
We also can define
$m_n(A):=\mbox{Tr}\,\rho A^n$ to be the $n^{\rm th}$ moment of $A$, and
$\kappa_2(A):=m_2(A)-m_1(A)^2$ to be the second cumulant, the variance,
uncertainty or dispersion of $A$ in the state $\rho$. 
von Neumann showed that there are no dispersion-free states.
Thus, quantum mechanics is intrinsically random.
Heisenberg's {\em uncertainty relation}, which is a theorem, not
a postulate, is the best-known facet of this:
\begin{theorem}
Let $A,B,C\in{\cal B}({\cal H})$ be such that $[A,B]:=AB-BA=C$; then in
any state $\rho$, we have $\kappa_2(A)\kappa_2(B)\geq m_1(C)^2/4$.
\end{theorem}
There is no uncertainty relation for commuting operators $A,B$, and
such observables are said to be
{\em compatible}. If $[A,B]\neq0$, we say that $A$ and $B$ are
complementary.

Segal has emphasised that the bounded observables in any quantum theory
should form the Hermitian part of a $C^*$-algebra with identity. This is
a complex vector space ${\cal A}$ with
\begin{enumerate}
\item a product $AB$ is defined for all $A,B\in{\cal A}$, which is
distributive and associative, but not necessarily commutative;
\item a conjugation $A\mapsto A^*$, which is complex-antilinear, is
specified;
\item ${\cal A}$ is provided with a norm $\|\bullet\|$ which
obeys Gelfand's condition
\begin{equation}
\|A^*A\|=\|A\|^2;
\end{equation}
\item ${\cal A}$ is complete in the topology given by this norm.
\end{enumerate}
This concept includes all the examples we have seen so far; the set
${\cal M}_n({\bf C})$, denoting $n\times n$ matrices, with matrix addition
and product, is a $C^*$-algebra. The $^*$ operation is Hermitian conjugate,
and the norm $\|A\|$ is the maximum eigenvalue of $|A|=(A^*A)^{1/2}$.
For any Hilbert space, ${\cal B}({\cal H})$ is also a $C^*$-algebra,
and more generally, so is any von Neumann algebra, which
can be defined as any weakly closed $^*$-subalgebra of ${\cal B}({\cal H})$
containing the identity. Another notable example
is the subset ${\cal C}(n)$ of ${\cal M}(n)$ consisting of real diagonal
matrices $A=\mbox{diag}\,(a_1,\ldots,a_n)$. This is clearly commutative, and
the diagonal elements are the eigenvalues. Thus, each $A\in{\cal C}$
determines uniquely a function $i\mapsto a_i,\;1\leq i\leq n$ from the
set $\Omega_n=(1,2,\ldots,n)$ to ${\bf R}$. Conversely, any random
variable $f$ on $\Omega_n$ defines a unique diagonal matrix diag$\,(f(1),
\ldots,f(n))$. So the classical observables on $\Omega_n$ can be described
as a special type of quantum mechanics, namely, the diagonal matrices.
Moreover, the interpretation in classical theory, of the values of the
random variables $f$ as possible observed values, coincides with the
quantum interpretation of the eigenvalues. Also, each $n\times n$
density matrix $\rho$ defines a unique probability measure $p$
on $\Omega_n$, by using the diagonal elements: $p(i)=\rho_{ii},\;1\leq 
i\leq n$. Clearly, a probability $p$ can define a density matrix by the
same formula, but there are other, non-diagonal density matrices giving
the same $p$. If all the observables are contained in ${\cal C}$, then
the off-diagonal elements of the density matrix are of no relevance,
and all the information on the state of the system is contained in $p$.
A concept that captures the essentials of this idea, removing redundant
description, is due to Segal. Given the algebra of observables, ${\cal A}$,
we say a {\em state} on ${\cal A}$ is a positive, normalised linear map
$\rho:{\cal A}\rightarrow{\bf C}$. Thus
\begin{enumerate}
\item $\rho$ is complex linear;
\item $\rho(I)=1$;
\item $\rho(A^*A)\geq0$ for all $A\in{\cal A}$.
\end{enumerate}
Naturally, two constructions that lead to the same map are said to define the
same state. We should note that we only need the expectations, i.e. the first
moments, of the observables, because ${\cal A}$ itself contains all
powers of $A$, and (as it is complete), also elements such as $e^{iA}$; so
if we know the state we know the characteristic function of every
observable, and so its distribution too.

More generally, the classical
measure theory $(\Omega,{\cal B},\mu)$, where $\mu$ is a positive
measure, can be written as a (commutative)
quantum theory by using the von Neumann algebra $L^\infty(\Omega,\mu)$
acting as multiplication operators on $L^2(\Omega,{\cal B},\mu)$;
its normal states correspond to (countably additive) probability
measures, which vanish on $\mu$-null sets. Indeed, given a state $\rho$
we can define the corresponding measure of a set $B\in{\cal B}$ as
$\rho(\chi_{_B})$. In this, sets $B$ and $B^\prime$ are indistinguishable
if the differ by a $\mu$-null set; we do not really need $\Omega$ itself,
but only the $\sigma$-tribe ${\cal B}$, modulo this equivalence.

The set of states of a $C^*$-algebra ${\cal A}$ forms a convex set, which
we shall call $\Sigma({\cal A})$ or just $\Sigma$. The convex sum
\begin{equation}
\rho=\lambda\rho_1+(1-\lambda)\rho_2,\hspace{.5in}\mbox{ where }0
<\lambda<1
\label{mixed}
\end{equation}
represents the random mixing of the states $\rho_1$ and $\rho_2$ with
weights $\lambda$ and $1-\lambda$. All expectations in the state
$\rho$ are then the same mixtures of the expectations in the states
$\rho_1$ and $\rho_2$. If $\rho_1\neq\rho_2$ we say that $\rho$ is a
{\em mixed state}. If $\rho$ cannot be written as a mixed state (so that in
any relation such as eq.~(\ref{mixed}) we must have $\rho_1=\rho_2$), the
we say that $\rho$ is a {\em pure state}. Every $C^*$-algebra possesses
many pure states. For the full matrix algebra ${\bf M}_n$, every
pure state is given by a unit ray $\{\psi\}$ in the Hilbert space
${\bf C}^n$, using the usual quantum-mechanical expression;
every density operator is a mixture of such.
This is an example of the Krein-Milman theorem, which says that a
weak$^*$-compact convex set in the dual of a Banach space is generated
by its extreme points. The representation of a mixed state as 
eq.~(\ref{mixed}) is, in general, not unique. For example, if ${\cal
H}={\bf C}^2$, the fully unpolarised state is $(1/2)I$, and this the
equal mixture of the pure states, the eigen-vectors of $J_3$, the spin
operator in the direction of quantisation, as well as the equal mixture
of the eigenstates of $J_1$, or any other spin direction.
This means that all statistical properties of the observables are the
same however the state was made up. We express this by saying that the
state-space in quantum probability is in general not a simplex: in a
simplex, any mixed state has only one decomposition into pure states. In
classical probability, in contrast, the state space $\Sigma(\Omega)$ is a
simplex. This is true in quantum probability only if ${\cal A}$ is
abelian. The density matrix contains all the information there is. Our
inability to
distinguish the history of how the state was made is due to the quantum
phenomenon of {\em coherent} sums of wave-functions.

There is an important connection between states and representations
of a $C^*$-algebra. A {\em representation} $\pi$ of ${\cal A}$ is
a $^*$-homomorphism from ${\cal A}$ into ${\cal B}({\cal H})$
for some Hilbert space ${\cal H}$. Thus, $\pi(A)$ is an operator on
${\cal H}$ and the map $\pi$ satisfies, for all $A,B\in{\cal A}$,
\begin{enumerate}
\item $\pi(\lambda A+B)=\lambda\pi(A)+\pi(B),\hspace{.3in}$ for all $\lambda
\in{\bf C}$ (linearity);
\item $\pi(A^*)=(\pi(A))^*\hspace{.3in}$(hermiticity).
\end{enumerate}
A representation is said to be faithful if $\pi(A)$ is
non-zero if $A\neq 0$. A state $\rho$ is said to be faithful if $\rho
(A^*A)=0$ only for $A=0$. To each state $\rho$ there is a representation
$\pi_\rho$, on a Hilbert space ${\cal H}_\rho$, and a unit vector $\psi_\rho
\in{\cal H}_\rho$, such that $\rho$ is vector state $\psi_\rho$; that is,
\begin{equation}
\rho(A)=\langle\psi_\rho,\pi_\rho(A)\psi_\rho\rangle,\;A\in{\cal A}.
\label{GNS}
\end{equation}
If the state $\rho$ is faithful, then so is the corresponding representation
$\pi_\rho$. Moreover, $\pi$ is irreducible if and only if $\rho$ is pure.

The proof of this theorem, which asserts the existence of ${\cal H}_\rho$
and the homomorphism $\pi_\rho$, follows the common mathematical trick:
we construct these objects out of the material at hand. Let us do this
when ${\cal A}$ has an identity and $\rho$ is faithful. We start with
the vector space ${\cal A}$ and provide it with the scalar product
\[\langle A,B\rangle:=\rho(A^*B).\]
The completion of this space is then taken to be ${\cal H}_\rho$. The
operator $\pi_\rho(A)$ is taken to be left-multiplication of ${\cal A}$
by $A$, thus: $\pi_\rho(A)B:=AB$. This defines $\pi_\rho(A)$ on the
dense set ${\cal A}\subseteq{\cal H}_\rho$, and can be shown to be bounded.
We take $\psi_\rho=I$, the identity in the algebra. One can then verify
that $({\cal H}_\rho,\pi_\rho,\psi_\rho)$ satisfy eq.~(\ref{GNS}).
A slightly more elaborate construction can be given if there is no identity
or the state is not faithful.
This realisation of the algebra is called the {\em GNS} construction, based on
$\rho$.

It took some time before it was understood that quantum theory is a
generalisation of probability, rather than a modification of the laws of
mechanics. This was not helped by the term quantum {\em mechanics}; more,
the Copenhagen interpretation is given in terms of probability, meaning
as understood at the time. Bohr has said \cite{Schilpp} that the
interpretation of microscopic measurements must be done in classical
terms, because the measuring instruments are large, and are therefore
described by classical laws. It is true, that the springs and cogs making
up a measuring instrument themselves obey classical laws; but this does not
mean that the {\em information} held on the instrument, in the numbers
indicated by the dials, obey classical statistics. If the instrument
faithfully measures an atomic observable, then the numbers indicated by the
dials should be analysed by quantum probability, however large the
instrument is.

We now present Gelfand's theorem, which shows that any
commutative quantum theory can be viewed as a classical probability
theory. We give a proof in finite dimensions.
\begin{theorem}
Given a commutative $^*$-algebra ${\cal C}$ of finite dimension, there
exists a (finite) space $\Omega$ and an algebraic $^*$-isomorphism $J$ from
${\cal C}$ onto ${\cal A}(\Omega)$, such that for any state $\rho$ on
${\cal C}$ there exists a probability $p$ on $\Omega$, such that for any
element $A\in{\cal C}$ we have
\begin{equation}
\rho(A)=E_p[J(A)].
\label{av}
\end{equation}
\end{theorem}
Proof\\
Since dim$\,{\cal C}=n<\infty$, the dimension of the dual space is the same.
There is a faithful state $\omega$ on ${\cal C}$; this could be for example
a mixture of a basis of the state-space with non-zero coefficients.
We can therefore construct a faithful realisation of ${\cal C}$ as a matrix
algebra. In this, the {\em GNS} construction,
the Hilbert space is built out of ${\cal C}$ and so is of
dimension $n$. A commutative collection of normal matrices can be simultaneously
diagonalised, so there is a basis in the Hilbert space such that each
element of ${\cal C}$ is a diagonal $n\times n$ matrix. Since exactly
$n$ of these diagonal matrices make up a linearly independent set, every
diagonal matrix appears. Every element of ${\cal C}$ is therefore a sum
of multiples of {\em units} $\{e_j\}$ of the algebra, satisfying $e_j^2=e_j$ and
$e_ie_j=0,\;i\neq j$. In the above matrix realisation, $e_j$ is the matrix
with $1$  on the diagonal in position $j$, and zero elsewhere.
Thus $A=\sum a_je_j$. So let $\Omega=\{e_j\}_{j=1,\ldots n}$, and let $JA$
be the function $JA(e_j)=a_j$. Then one verifies that $J$ is an algebraic
$^*$-isomorphism. To the state $\rho$ we associate the probability
$p(e_j)=\rho(e_j)$, and see easily that eq.~(\ref{av})
holds.\hspace{\fill}$\Box$

In this proof, instead of identifying $\Omega$ with the collection of
elements $e_j$ in the algebra, we could have taken the dual, and
identified $\Omega$
with the set of {\em characters} on ${\cal C}$. This is the set of
multiplicative states, that is, states $\omega$ obeying $\omega(AB)
=\omega(A)\omega(B)$ for all $A,B\in{\cal C}$. The set of
characters of a $C^*$-algebra is called
its {\em spectrum}. Our proof shows that there are exactly $n$ of these, 
defined by $\omega_j(e_k)=\delta_{jk}$. Putting $A=B$ we see that any
character is dispersion-free. This is why the spectrum is taken by Gelfand
to be the definition of $\Omega$ in the infinite-dimensional case:
\begin{theorem}
Let ${\cal C}$ be a commutative $C^*$-algebra with identity. Then the set of
characters can be given a topology so as to form
a compact Hausdorff space $\Omega$ such that ${\cal C}$ is
$C^*$-isomorphic to $C(\Omega)$, and every state on ${\cal C}$ corresponds
to a finitely additive measure on $\Omega$ (with the Borel tribe).
\end{theorem}
 
Bohm asked whether the observed statistics, agreeing with
experiment, can be obtained from a larger, more complicated classical
theory. This is the idea behind the attempts to introduce hidden
variables. This is certainly true of the statistics of any fixed complete
commuting set of observables; for they form an abelian algebra, and so can
be represented by the classical statistics of multiplication operators
on a sample space (the spectrum of the algebra). Obviously
the full non-abelian algebra cannot be a subalgebra of an abelian algebra,
so the way hidden variables are introduced must be more elaborate
than extending the algebra by adding them. However, the deep
result of J. S. Bell shows (if the dimension is 4 or higher) that the
full set of statistics predicted by quantum theory cannot be got
from {\em any} underlying classical theory. In the quantum model
of two spin-half systems, Bell constructs a sum of four correlations
which in a certain state is equal to $2\surd 2$, a factor $\surd 2$
larger than the greatest value allowed in any classical theory.

Let us follow \cite{Landau,Kochen}. Let $P,Q$
be complementary projections, and also let $P^\prime,Q^\prime$ be
complementary projections, while $P$ is compatible with $P^\prime$
and with $Q^\prime$, and $Q$ is compatible with $P^\prime$ and
$Q^\prime$. Define $a=2P-I$, $b=2Q-I$, and similarly for $a^\prime$
and $b^\prime$. For any state $\rho$ define $R$ by
\[R:=\rho(aa^\prime+ab^\prime+bb^\prime-ba^\prime)=\rho(C)\]
where $C=a(a^\prime+b^\prime)+b(b^\prime-a^\prime)$. Then $a^2=b^2=a^{\prime
2}=b^{\prime 2}=1$, so
\begin{equation}
C^2=4+[a,b][a^\prime,b^\prime]=4+16[P,Q][P^\prime,Q^\prime].
\label{Bell}
\end{equation}
Since $\|a\|=\|b\|=\|a^\prime\|=\|b^\prime\|=1$, it follows that
\[\|[a,b][a^\prime,b^\prime]\|\leq 4,\]
so $C^2\leq 8$ and $|R|^2=|\rho(C)|^2\leq\rho(C^2)\leq 8$. So in quantum
theory, $|R|\leq 2\surd 2$.
If there is a joint probability space on which we can describe $a,\ldots,
b^\prime$ by the r. v. $f,\ldots g^\prime$ taking the values
$\pm 1$, and a measure $p$ on it, then $R=E_p[h]$ where
\[h=f(f^\prime+g^\prime)
+g(g^\prime-f^\prime).\]
Then these r. v. commute, so eq.~(\ref{Bell}) becomes $h^2=4$, and
\[|R^2|=E_p[h]^2\leq E_p[h^2]=4.\]
So $|R|\leq 2$, (Bell's inequality). Bell showed that the entangled
states of the Bohm-EPR set-up give a $\rho$ such that $R=2\surd 2$,
violating this. Thus no description by classical probability is possible.

The famous Aspect experiment tested Bell's inequalities.
This involves observing a system (in a pure entangled state) in a long run
of measurements; the correlations singled out by Bell,
between several compatible pairs of spin observables, were measured.
The experiments showed that $R$ was just less than $2\surd 2$, in
agreement with the quantum predictions.

The upshot is that in quantum probability there is no sample space;
we have the $C^*$-algebra ${\cal A}$, and this plays the r\^{o}le of
the space of
bounded functions.

Let us now examine Bohm's claim that there is a hidden
assumption in Bell's proof, that of `locality'. It is now generally
agreed that the term `local', referring to the space-localisation,
is not the best, and that `non-contextual' is a better term; namely
that the choice of random variable $f$ assigned to represent a certain
observable $a$ which is being measured, does not depend on which of the
other observables, $a^\prime$ or $b^\prime$, is being measured at the
same time. This is now called a non-contextual assignment.
Bohm suggested that we should allow a contextual choice of
assignment of random variable, so that the r.v. representing the observable
$a$ when $a^\prime$ is also measured is not the same as the choice of r.v.
for $a$ when it is measured with $b^\prime$.
The two choices will, however, have the
same distribution. It should be said straight away that this idea is
contrary to the practice of probabilists, who would expect there to be
a unique random variable representing an observable. It also goes against
the definition of `element of reality' of {\em EPR} as extended by us
to the random case. The quantum version does not suffer from this unreality,
since the mathematical object assigned to the observable, the Hermitian
matrix, does not depend on the context, i.e. is local in Bohm's language.

Bohm's idea leads to a theory with very few rules.
However there are some restrictions, since the choice must
be done so that {\em all}
statistical measurements of compatible observables (means, correlations,
third moments etc) of the
model can be arranged to give the same answers as the quantum theory.
This is achieved as follows. Let $a,a^\prime$
be compatible, generating a commutative
$C^*$-algebra, ${\cal C}$ and let $\rho$ be a state on the full algebra
${\cal A}$. By restriction, $\rho$ defines a state on ${\cal C}$.
By Gelfand's isomorphism,
we can construct a space $\Omega$, the spectrum
of ${\cal C}$, and a measure $\mu$ on it, such that $a,a^\prime$ can be
represented as multiplication operators on ${\cal C}(\Omega)$, so
they are random variables, $f,g$. The joint probability
distribution of $f,g$ is the same as that of the (diagonal) matrices
$a,a^\prime$ in the state $\rho$.
On the other hand, $\Omega,\mu$ depends on the  set $a,a^\prime$. Let
us record this by denoting this Gelfand representation by $\Omega_{a,
a^\prime},\mu_{a,a^\prime}$. If we
measure $a$ and $b^\prime$, and proceed as Bohm suggests, then we get a
different space $\Omega_{a,b^\prime}$, the spectrum of a different algebra
${\cal C}_{a,b^\prime}$, say. The state $\rho$ leads to a different measure
$\mu_{a,b^\prime}$.
The r.v. assigned to $a$ cannot be $f$ this time;
it must a function on $\Omega_{a,b^\prime}$, a different space;
it has the same distribution in $\mu_{a,b^\prime}$ as the $f$ had in
$\mu_{a,a^\prime}$. In this set-up, there is no obvious definition of
$a^\prime+b^\prime$, as they are not functions on the same space.
This problem does not arise in the quantum formulation:
there is an underlying $C^*$-algebra, in which we can add the operators.

Bohm's suggestion might be said to be an
interpretation of quantum mechanics in terms of classical probability
\cite{Garden}.
However, the construction is not a probability theory in the sense
of Kolmogorov, as there is no single sample space; the theory is
preKolmogorovian, in the tradition of the frequentist school.
One can generalise the frequentist point of view, and specify
that certain collections of observations are
compatible, and others are not; then we can by observation construct
the joint probabilities of each compatible set, and have no need of
the sample space (the space of joint values). A different compatible set
need have no analytic relation to the first, even though it contain
common observables. Bell's inequality need not hold, but then
neither need the quantum version, which is $\surd 2$ times
more generous. It is a feeble theory, not much more that data
collection, and has no predictive power. Mere data give us no
more than mere data.

Another variant of quantum mechanics, a new form of algebra
called `quantum logic', was developed in
\cite{Birkhoff2}. New rules by which propositions can be
manipulated are given. This was worked on later by Jauch and coworkers \cite{Jauch},
culminating in Piron's thesis. This says that the propositions
form a lattice isomorphic to the
lattice of subspaces in a Hilbert space (but not necessarily over the
complex field). Apart from this result, quantum logic has not been very
successful, and it is more productive to keep to
classical logic, but to generalise the concept of probability algebra
from commutative to non-commutative. Another alternative to
quantum probability is stochastic
mechanics, founded but now abandonned by Nelson \cite{Nelson4}
as not being correct physics. Thus Segal's approach is the one we adopt here.
It is well explained in \cite{Emch,Haag,Horuzhy}.

Quantum theory has its version of estimation theory \cite{Holevo,Ohya}.
In finite dimensions, the method of maximum likelihood is to find the
density
matrix $\rho$ that maximises the entropy, subject to given values for
the means, $\{\eta_i\}$ of observables in the subspace of hermitian operators
spanned by a named list $\{X_1,\ldots,X_n\}$ of {\em slow variables}.
So $\eta_i=\mbox{Tr}[\rho X_i]$. It is well known
that the answer is the Gibbs state
\begin{equation}
\rho=Z^{-1}\exp(-H)=Z^{-1}\exp-[\xi^1X_1+\ldots+\xi^nX_n],\hspace{.2in}
Z=\mbox{Tr}[\exp(-H)].
\end{equation}
Again, $\log Z$ is strictly convex, and its Hessian gives a Riemannian
metric on the manifold ${\cal M}$ of all faithful density operators
\cite{Ingarden4,Chentsov,Petz2}. In this case we get the
Kubo-Mori-Bogoliubov metric; in terms of the centred variables
$\hat{X}_i:=X_i-\eta_i$,
the metric is
\begin{equation}
g(\hat{X}_i,\hat{X}_j)=\mbox{Tr}\left[\int_0^1\rho^\lambda\hat{X}_i\rho^
{1-\lambda}\hat{X}_jd\lambda\right]
\end{equation}
This is the closest point on ${\cal M}$ to any state with the given
means, where `distance' is measured by the relative entropy $S(\rho|
\rho^\prime):=\mbox{Tr}\,\rho[\log\rho-\log\rho]$.
Again, the $\xi^j$ are uniquely determined by
the measured means $\eta_i$.

\section{Kolmogorov and Ito}
A {\em stochastic process} over a set $T$ is a family $\{X_t,t\in T\}$
of random variables on a measure space $(\Omega,{\cal B},\mu)$.
We might have $T=\{0,1,2\ldots\}$, or $T={\bf R}^+$, when we interpret
$t$ as time. From the frequentist point of view, we can observe
$X_{t_1},X_{t_2},\ldots X_{t_N}$ at finitely many points of time.
In this way, we can test any a model as to what the joint distribution
of these r.v. is.

Kolmogorov's existence theorem says that a family of joint (cumulative)
distributions
$F_{1,2\ldots n}(x_1,\ldots,x_n)$, given for all finite subsets of $T$, is
the set of joint distributions of a stochastic process over $T$ if and only
if the {\em consistency conditions} hold. Thus, (the hatted variable is omitted):
\begin{enumerate}
\item For any permutation $\pi$ of $(1,2,\ldots,n)$, we have
\[F_{1,\ldots n}(x_1,\ldots,x_n)=F_{\pi(1),\ldots,\pi(n)}(x_{\pi(1)},
\ldots,x_{\pi(n)})\]
\item For any $j$, we have
\[ F_{1,\ldots,n}(x_1,\ldots,x_j=\infty,x_{j+1},\ldots,x_n)=F_{1,\ldots,
\hat{j},\ldots,n}(x_1,\ldots,\hat{x_j},\ldots,x_n).\]
\end{enumerate}

If these hold, he shows that the sample space $\Omega$ may be taken to
be ${\bf R}^T$, an enormous space (of all functions $\omega$ of $t$); the
r. v. $X_t$ is then the function $X_t(\omega)=\omega(t)$.
He proved the existence of
a measure $\mu$, which reproduces the given joint distributions; the
$\sigma$-tribe ${\cal B}$ has the following structure. 
Let ${\cal B}_t$ be the smallest $\sigma$-tribe such that all $X_s$, for
$s\leq t$, are measurable; then this is an increasing family of
$\sigma$-tribes, called a filtration. Then ${\cal B}$ is the smallest
$\sigma$-tribe containing all the ${\cal B}_t$.

Apply this to the Brownian paths, and the
measures defined by a finite set of gates as in the last section;
this proves
that there is a probability theory underlying the finite joint
distributions. However, it does not prove Wiener's theorem, in that the
sample space obtained by the Kolmogorov construction is the huge set of all
functions of time. It is then a hard problem to show that the subset
of continuous functions has measure 1. This fact is very important
for specialists in Brownian motion, but is not a general feature of
processes covered by Kolmogorov's theorem, and is not needed to construct
the usual $L^p$ spaces of functional analysis.
Without Wiener's version
we lose the power of the path-wise methods, and also lots of intuition.
The modern method is to get the cow off the ice using Kolmogorov, and
supplement it with further estimates, on tightness and radonifying maps,
if we need to find smaller carrier spaces for the measure
\cite{Gross,Schwartz}.
After Kolmogorov's treatise, the subject could develop `in the usual
professional mathematical way', to use Segal's phrase. That is, theorems
could be stated and proved, and then sharpened. The most important of these
were the laws of large numbers, the zero-one laws, the central limit
theorems,
the theory of large deviations, the classification of all processes with
independent increments, martingales, and stochastic integration.

The conditional
expectation $E_t:=E[\bullet|{\cal B}_t]$ takes a random variable in
$L^2(\Omega,{\cal B},\mu)$ into one in $L^2(\Omega,{\cal B}_t,\mu)$;
since it is the identity on the latter space, and is Hermitian,
it must be the orthogonal projection onto $L^2(\Omega,{\cal B}_t,\mu)$.
None of these ideas depends on which version of the sample space we have.

The concept of conditional expectation can be extended to integrable
r.v., thus:
\begin{definition}
Let $(\Omega,{\cal B},\mu)$ be a probability space, and let ${\cal B}_0$ be
a sub-$\sigma$-tribe of ${\cal B}$. Let $X$ be a random variable with
$E[X]<\infty$. Then there exists a ${\cal B}_0$-measurable r.v. $Y$,
written $E[X|{\cal B}_0]$, such that for each set $B\in{\cal B}_0$ we have
\begin{equation}
\int_BY\,d\mu=\int_BX\,d\mu
\end{equation}
Further, if $\hat{Y}$ is another r.v. with these properties, then
$\hat{Y}=Y$ almost everywhere.
\end{definition}
See \cite{Williams} for a proof, and other things.

A martingale is a stochastic process $X_t$ on $(\Omega,{\cal B}_{t\geq 0}
,\mu)$ such that $X_t$ is integrable, and
\begin{equation}
E[X_t|X_s]=E[X_s]\mbox{ for all }t\geq s.
\label{martingale}
\end{equation}

A martingale is a fair game.
For example, consider the independent tosses of a fair coin,
and let $X_n=H_n-T_n$, where $H_n$ is the number of heads and $T_n$ is
the number of tails at the $n^{\rm th}$ toss. Let $S_N=\sum_{j=1}^NX_j$.
Then $S_N$ is a martingale \cite{Grimmett}, p 202.

There are four concepts of convergence of a sequence $\{X_n\}$
to $X$ in the space of random variables on a probability space $(\Omega,
{\cal B},\mu)$.
\begin{enumerate}
\item We say $X_n\rightarrow X$ {\em almost surely} (or, almost
everywhere) if
\[\mu\{\omega:X_n(\omega)\rightarrow X(\omega)\}=1.\]
\item We say $X_n\rightarrow X\mbox{ in }\|\bullet\|_r$ if
\[\|X_n-X\|_r\rightarrow0\mbox{ as }n\rightarrow\infty.\]
\item We say that $X_n\rightarrow X$ in probability if
\[\mu\{\omega:|X_n(\omega)-X(\omega)|>\epsilon\}\rightarrow0 \mbox{ as }
n\rightarrow\infty\;\;\mbox{for all }\epsilon>0.\]
\item We say $X_n\rightarrow X$ in law if
\[\mu\{\omega:X_n\leq x\}\rightarrow\mu\{\omega:X\leq x\}\mbox{ for all }
x\mbox{ at which }F_X(x):=\mu\{X\leq x\}\]
is continuous.
\end{enumerate}
These concepts are not equivalent; (1) and (2) are not comparable, but
(1) or (2) imply (3), which
implies (4) \cite{Grimmett}. Convergence in law can be
related to convergence of the characteristic functions of $X_n$ to that of
$X$; we see that if $X_n$ converges to $X$ in law implies that $X_n$
also converges to $Y$ in law if $Y$ has the same distribution as $X$.
This shows that convergence in law in a very feeble concept.
The four concepts of convergence do not depend
on the version of sample space adopted, and so are the same whether we use
Wiener space or Kolmogorov's abstract construction. 

For a given $\mu$, we can complete the $\sigma$-tribe ${\cal B}$ to include
all subsets of sets of $\mu$-measure zero; then the events that can happen
are described by the quotient $\sigma$-tribe, in which events which differ
by a set of measure zero are identified. This idea is not wise when we are
interested in measures with different sets of zero measure, as happens when
we condition a Wiener path to pass through a given point. The Dirac measure
on ${\bf R}$ is a simple example of the trouble we get into. If two
measures $\mu_1,\mu_2$ have the same sets of zero measure in $(\Omega,{\cal
B})$, we say that they are equivalent. If $\mu_1(B)=0$ whenever $\mu_2(B)=0$
we say that $\mu_1$ is absolutely continuous relative to $\mu_2$; in that
case there exists a function $w\in L^1(\Omega,{\cal B},\mu_2)$ such that
$\mu_1(B)=\int_B\rho(\omega)d\mu_2$ for all $B\in{\cal B}$. We write
$w=d\mu_1/d\mu_2$, the Radon-Nikodym derivative. This is the abstract
version of eq.~(\ref{change}). We shall be interested in other measures,
singular relative to a given one. Then the best formalism is to start with
an abelian $C^*$-algebra ${\cal A}$ and consider its states.

Estimation is assisted by the law of large numbers.
Let $X$ be a random variable on a probability space, whose mean $\eta$ we
wish to find, making use of a random experiment which is believed to be
well modelled by $X$. We set up a sequence of independent copies $X_n$
of $X$, and
consider the stochastic process $\{X_n\}$ on e.g. the probability space
constructed by the theorem of Kolmogorov. The {\em strong law of
large numbers} says that if $E(X)=\eta$ and $E(X^2)<\infty$, then putting
$S_n=\sum_{j=1}^N S_j$ we have
\[S_N/N\rightarrow \eta I \mbox{ in }\|\bullet
\|_2.\]
If $E|X|<\infty$, we get almost sure convergence. These are necessary and
sufficient conditions.
Weaker conditions ensure that the sum converges in probability
\cite{Grimmett}. This is called the weak law.
Note that the meaning of convergence uses the measure on the Kolmogorov space,
so for almost all sequences, randomly chosen, we get convergence to the mean.
It does not say how fast the convergence is.
For example if $X_n$ is the number of heads minus the number of tails,
at the $n^{\rm th}$ toss of a fair coin, then $S_N$ is the number of heads in $N$
tosses minus the number of tails, and $S_N/N$ converges almost surely to
zero. If we know that $S_m\neq 0$ after $m$ results, the law does not say
that the bias evens out in the long run.
$S_n$ is a martingale, and its expected value for $N>m$ is its present
value $S_m$. It is $S_N/N$, which converges; the bias at time
$m$ gets divided by $N$, and so goes away for large $N$.

Another famous limit law is the central limit theorem; if the standard
deviation $X$ is 1 and the mean is zero, then one can show that
\[S_n/(\surd n)\rightarrow N(0,1)\mbox{ in law}.\]
Versions of this were known to Bernouilli and Gauss, if we assume that the
moment-generating function exists. It explains the ubiquity of the normal
distribution; many random processes are the sums of small and independent
random things, and so tend to be Gaussian.
The theory of large deviations tells us something about the rate of
convergence of $S_n$; this stuff is deeper \cite{Varadhan,Donsker,Lewis,Stroock2}.
There is also a large body of work on sums of nearly independent random
variables, and also on the cases where the variances are not all equal.

Doob proved that martingales often converge; e. g.,
\begin{theorem}
If $\{S_n\}$ is a martingale with $E(S_n^2)<M<\infty$ for some $M$ and all
$n$, then there exists a random variable $S$ such that $S_n\rightarrow S$
almost surely.
\end{theorem}

Consider now a process $(X_t,\Omega,{\cal B}_{t\geq 0},\mu)$ in continuous
time with independent increments; that is, $X_t-X_s$ is independent of
$X_r$ for $r<s<t$. Since we can write $X_t-X_s$ as the sum of more and more
independent differences, we might expect that $X_t-X_s$ must be Gaussian,
by the central limit theorem. However, this is not the case since
the distributions of the difference $X_t-X_s$ might change as the interval
is made smaller.
This question led Levy to characterise all processes that are
stationary and have independent increments. This can be done by showing
that the characteristic function
\[ C(\lambda):=E[\exp{i(X_t-X_s)\lambda}]\]
should not only be of positive type, but so should any fractional power.
Such a function is called {\em infinitely divisible}, and so is the
corresponding random variable. The necessity is easy to see; we can write
$X_t-X_s$ as the sum of $N$ identical and independent random
variables, namely, the increments for time intervals $(t-s)/N$; then the
characteristic function of this sum is the product of the $N$
characteristic functions (which are all equal, by stationarity) of
these increments, and so C has an $N^{\rm th}$ root that is of positive type.
This condition is also sufficient, to which we shall return.
The characteristic function of the Gaussian is infinitely divisible, and
so is that for the Poisson distribution. This means that Gaussian and Poisson
processes with independent increments exist. Levy found that by mixing these
he got some new processes (Levy processes), and he found
the most general form of the characteristic function, which is
\begin{equation}
\log C(\lambda)=-a\lambda^2+ib\lambda+\int d\sigma(\alpha)
[e^{i\alpha\lambda}-(1+i\alpha\lambda)Z(\alpha)]
\label{Levy}
\end{equation}
where $a\geq0$, $b$ is real, $d\sigma(\alpha)\geq0$ obeys $\int_{-1}^1
\alpha^2d\sigma(\alpha)<\infty$. There are some
further conditions on the weight $d\sigma$ at
infinity \cite{GelfandV}. If $\sigma=0$ we get the Gaussian, and if
$\sigma$ has a discontinuity, we get a Poisson process. These can be
understood in terms of Hilbert space cohomology of ${\bf R}$, as in
\S(5).

During this period, physicists and engineers studied stochastic differential
equations, similar to the Langevin equation. Often the random force was
chosen to be the derivative of Brownian motion, called white noise. Since
$B_t$ is at best continuous, this work lacked rigour, and remained
poorly defined even after appeal to Dirac's generalised concept of
function. This sorry state of affairs was cleared up by Ito.

Let $W(t)$ be Brownian motion starting at zero. At first sight, an equation
for an unknown $X(t)$ similar to the Langevin equation, of the form
eq.~(\ref{Langevin})
\[\frac{dX_t}{dt}=a(t)+b\frac{dW_t}{dt}, \mbox{ for almost all }\omega\]
makes no sense, since for almost all $\omega\in\Omega$, $W(t)$ is not
differentiable. The equation does make sense if written in the form:
find a family of random variables, $\{X(t)\}$, such that
for a given initial random variable $X(0)$, the r. v.
$X(t)-X(0)-\int_0^ta(s)ds$ 
is the known r.v. $W_t$ for almost all $\omega$. This does not prove that
there is such a process, but it is does make sense. For the more general
case when $a,b$ depend on the unknown $X(t)$, the integral form is
\begin{equation}
X(t)-X(0)=\int_0^ta(s,X(s))ds+\int_0^tb(s,X(s))dW(s).
\label{Itoequation}
\end{equation}
The last expression, called a stochastic integral, looks like a Stieltjes
integral, but the needed condition of bounded variation on $W(s)$ do not
hold.
Solve the equation by iteration (Picard's method); we see that
at each stage, the approximation to $X(t)$ is a function of $W(s)$ only for
$s\leq t$. So it would appear that we need only
give a meaning to the stochastic integral for the cases where $X(t)$
is a function of $W(s)$ for $s\leq t$, and so the same holds for $b(t,X(t))$.
This can be neatly put in terms of the filtration ${\cal B}_t$ generated by
the Wiener process: for all $t\geq 0$, $X(t)$ and so $b(t,X(t))$
is measurable relative to the $\sigma$-tribe
${\cal B}_t$. This makes sense physically; it says that
we can know the present configuration $X(t)$ if we know the initial
configuration $X(0)$ and the outcomes of all the randomness, $W_s,\;s\leq
t$, so far. A random function of time, $f$ is said to be {\em adapted to
the filtration} ${\cal B}_t$ if $f(t)$ is ${\cal B}_t$-measurable for all
$t\geq0$. 

Let $f(t)$ be an adapted process in the time interval $0\leq t\leq T$,
which is {\em simple}: that is there is a finite partition $0=t_0,t_1,
\ldots<t_n=T$ such that $f(t)=f_j$ for $t_{j-1}\leq t<t_j$ for all
integers $j\in(1,\ldots,N)$. Here, $f_j$ is a random variable independent
of time, and equality of random variables means almost everywhere.
Following Ito, we can define the stochastic integral of an adapted simple
function $f$ to be the random variable
\begin{equation}
\int_0^Tf(t)dW_t:=\sum_jf_j(W_{t_{j+1}}-W_{t_j}).
\end{equation}
Note that the increment $dW$ is in the future of the random variable
$f_j$ that multiplies it. The mapping, $f\mapsto\int_0^Tf(t)dW_t$ takes
the linear space of simple adapted functions into the space
of random variables, and is clearly a linear map. The brilliant
remark of Ito is then that the following identity, called Ito's isometry,
holds:
\begin{equation}
E[|\int_0^Tf(t)dW_t|^2]=\int_0^TE[|f(t)|^2]dt
\label{Ito}
\end{equation}
Proof:
\begin{eqnarray*}
E[|\int_0^Tf(t)dW_t|^2]&=&\sum_i\sum_jE[f_i(W_{t_{i+1}}-W_{t_i})f_j(W_{t_{
j+1}}-W_{t_j})]\\
&=&\sum_iE[|f_i|^2(W_{t_{i+1}}-W_{t_i})^2\\
&+&2\sum_{i<j}f_if_j(W_{t_{i+1}}-
W_{t_i})(W_{t_{j+1}}-W_{t_j})].
\end{eqnarray*}
Now, the future increment $W_{t_{i+1}}-W_{t_i}$ is independent of $f_i$,
which is adapted, i.e. a function of earlier $W(s)$. So the expectation
value in the first term factorises:
\begin{eqnarray*}
E[|f_i|^2(W_{t_{i+1}}-W_{t_i})^2]&=&E[|f_i|^2]E[(W_{t{i+1}}-W_{t_i})^2]\\
&=&E[f_i|^2](t_{i+1}-t_i),
\end{eqnarray*}
by the property of Brownian motion. This gives the desired term in
eq.~(\ref{Ito}). It remains to show that the remaining double sum vanishes.
This is true, because the factor $(W_{t_{j+1}}-W_{t_j})$ for $j>i$ is
independent of the remaining factors $f_if_j(W_{t_{i+1}}-W_{t_i})$ and so
the expectation of the product is the product of the expectations;
but the expectation of the future increment of $W_t$
is zero.\hspace{\fill}$\Box$

Ito's isometry is a mapping from the set of simple adapted processes
to random variables; by a simple theorem of normed spaces, it
can be extended by continuity to a linear isometry
(unitary transformation) between the completions of both sides in the
norms given. The completion of simple functions in the norm
\begin{equation}
\|f\|^2=\int_0^TE[|f(t)|^2]dt
\end{equation}
is the space of processes such that $E[|f|^2]$ is Lebesgue integrable; so
Ito can define the stochastic, or Ito integral, of all processes with this
property; it is the limit in this norm of simple adapted processes
approximating it. Naturally, we must prove that the adapted simple processes
are $L^2$-dense in the square-integrable adapted processes; this is not
difficult, since the projection $E_t$ is a bounded operator and maps onto
the space of ${\cal B}_t$-adapted square-integrable processes.

We can now give a meaning to the question, do there exist solutions to the
stochastic differential equation
\begin{equation}
\frac{dX_t}{dt}=a(X_t,t)+b(X_t,t)\frac{dW_t}{dt}?
\label{sde}
\end{equation}
We say the a process $X_t$ satisfies this equation if, on substituting
$X_s$ in the integrals in eq.~(\ref{Itoequation}) we get back $X_t-X_0$.
 
For a wide class of functions $a$ and $b$ of two variables we can then
get a convergent iterated approximation, the Picard series, which converges
to a process $X_t$ obeying the (integral form of) the stochastic
differential equation. This holds for example if $a(x,y)$ is
uniformly Lipschitz in $y$ in a region, and $b(x,y)$ is uniformly elliptic
in $y$ and measurable in $x,y$.
This result can be improved and generalised, so
that vector-valued stochastic processes can be studied, and the noise
can be of a much more general martingale than $W_t$.
This can be reworded as a `martingale problem' \cite{Stroock}.

The converse to Ito integration should be a form of differentiation:
it is called (Ito) stochastic differentiation; we may say that the process
$f(W_t,t)$ is the stochastic derivative of $\int_0^t f(W_s,s)\,dW_s$.
The Ito integral is always a martingale, and every martingale is
a stochastic integral, and so has a stochastic derivative, namely the
integrand in its representation as an Ito integral. One can show that
this is unique. It is interesting to form the repeated stochastic integrals
\begin{eqnarray*}
W_t&=&\int_0^tdW_s\\
:W_t^2:=W_t^2-t&=&2\int_0^tW_sdW_s\\
:W_t^3:&=&3\int_0^t:W_s^2:dW_s\\
\ldots& &\ldots
\end{eqnarray*}
in which the Wick ordered (Hermite polynomials) occurring in the Wiener chaos
are the successive stochastic integrals. They are all contained
in the exponential martingale $e^{\lambda X_t-\frac{1}{2}\lambda^2t}$.
The second one illustrates the Doob-Meyer decomposition: $W_t^2$
is a submartingale, and is written as the sum of a martingale, $:W_t^2:$,
and an increasing function, $t$ of bounded variation.

Manipulation of stochastic integration can be summarised
by the Ito multiplication table: keep all differentials in $dt$
up to first order, using $dt.dW=0$ and $dW.dW=dt$. From this, we can get
the important relation between a certain parabolic partial differential
equations known as Kolmogorov's forward equation, and the corresponding 
stochastic differential equation. Suppose that
$X_t$ satisfies the stochastic differential equation eq.~(\ref{sde}),
with initial r.v. equal to $X_0$, which has law $p(x)$.
Let $p(x,t)$ be the law of $X_t$; then it can be shown that
$p(x,t)$ is smooth and satisfies the parabolic equation
\begin{equation}
\frac{\partial p}{\partial t}=\frac{1}{2}\frac{\partial}{\partial x}\left
(b(x,t)^2\frac{\partial p}{\partial x}\right)+\frac{\partial}{\partial x}
(a(x,t)p),
\label{63}
\end{equation}
with initial condition $p(x,0)=p(x)$.
To see why this is, we note that if $f(x)$ is any smooth function,
we can apply Ito's lemma to the random process $f(X_t)$. We
recover $\int p(x,t)f(x)\,dx$ as $E[f(X_t)|X_t=x]$.
We now expand $f(X_{t+dt})$ in a Taylor series about $X_t$ up to
second order in $dW$:
\begin{equation}
f(X_{t+dt})=f(X_t)+\frac{\partial f}{\partial x}dX+\frac{1}{2}
\frac{\partial^2f}{\partial x^2}(dX)^2.
\label{eq}
\end{equation}
Eq.~(\ref{sde}) tells us that $(dX_t)^2=b^2\,dt$ and $dX=a\,dt +b\,dW$.
Here, $dW$ is the forward difference. Then
the expectation  vanishes: $E[f^\prime b\,dW|X_t-x]=E[f^\prime b|X_t=x]
E[dW|X_t=x]=0$,
since $dW$ is independent of $f^\prime b$ at time $t$,
and has zero expectation. So, taking the conditional
expectation of eq.~(\ref{eq}),
\begin{equation}
E[f(X_{t+dt}-f(X_t)|X_t=x]=E[\frac{\partial f}{\partial x}a]\,dt+
\frac{1}{2}E[b^2f^{\prime\prime}|X_t=x]\,dt.
\end{equation}
Since $a,b,f,f^\prime,f^{\prime\prime}$ are functions of $X_t,t$
they become sure functions, evaluated at $x$
under the conditioning; thus we get the equation for the increment
$f(x,t+dt):=E[f(X_{t+dt}|X_t=x]$:
\[(f(x,t+dt)-f(x))/dt:={\cal L}f=
(1/2)b(x,t)^2f^{\prime\prime}+a(x,t)f^\prime.\]
This is Kolmogorov's {\em backward} equation, which applies to the
dynamics of the process. To get the dynamics of the probability density,
we take the dual operator ${\cal L}^*$, defined by 
\[ \int p(x,t){\cal L}f(x,t)\,dx=\int{\cal L}^*p(x,t)f(x,t)dx\]
which on integration by parts, and discarding the boundary term at $\infty$
gives
\[{\cal L}^*f:=\frac{1}{2}\frac{\partial}{\partial x}\left(b(x,t)^2\frac
{\partial}{\partial x}f\right)+\frac{\partial}{\partial x}\left(a(x,t)f
\right).\]
Since $f$ was arbitrary, we see that $p(x,t)$ satisfies the forward
equation in the weak sense (after smoothing with a test-function $f$).
It is known from the theory of elliptic regularity, that any weak
solution is a strong solution. If $a$ and $b$ are constants, we arrive at
the Smoluchowski equation, and the continuum version of (\ref{KBE}):
\[p(x,t)=E[p(X_t,0)|X(0)=x].\]
This representation for the solution of the pde gives an immediate proof
that the solution remains non-negative if the initial condition is
non-negative, since $p(X_t,0)\geq 0$; also, one sees that the time-evolution must be a contraction
in the $L^\infty$-norm, and the $L^2$-norm as the conditional expectation
is a projection.

Sometimes, we can rewrite the solution $X_t$ in terms of time-translation
$\omega\mapsto\omega T_t$ if we modify the measure \cite{Williams2}.
Suppose $\mu^\prime$
is absolutely continuous relative to $\mu$. Then there exists an adapted
process $u(t)$ in $(\Omega,{\cal B},\mu)$ such that
\begin{equation}
dX_t=dW_t+u(t)dt,\hspace{.3in}X_0=0,
\label{CM}
\end{equation}
has a weak solution $X_t$ whose law is the same as $Y_t(\omega):=\omega(t)$
as a r. v. on $(\Omega,{\cal B},\mu^\prime)$. Then the Radon-Nikodym
derivative is
\begin{equation}
d\mu^\prime/d\mu^=\exp\left[\int_0^tu(s)dX_s-(1/2)\int_0^t\|u(s)\|^2ds\right].
\label{Girsanov}
\end{equation}
Conversely, if $u$ is such that the r. h. s. of (\ref{Girsanov}) has Wiener
expectation 1, (as will happen if $u$ is bounded), then there exists
an absolutely continuous measure $\mu^\prime$ given by (\ref{Girsanov}),
such that $T^*_t$ on $(\Omega,{\cal B},\mu^\prime)$ produces a weak solution
to (\ref{CM}).
This is the Girsanov-Cameron-Martin theorem.

This change of measure is closely linked to the change of
ground state in the corresponding quantum theory, when an interaction is
introduced. We see this in the Feynman-Kac formula, below.

One can, using similar methods, integrate adapted functions relative to
$dM$, where $M$ is any martingale. The stochastic integral has other
variants, such as the Stratonovitch
version \cite{Wiener2,Nelson4}; one can also integrate non-adapted
processes, subject to
other conditions (Skorokhod), or use another noise which is not quite
a martingale \cite{McShane,Barnett3}.
The Ito version has an interesting interpretation in mathematical finance.
Suppose that the price of an asset is a random process $S_t$, and it
obeys the Ito equation
\[dS_t=a(S_t,t)dt+b(S_t,t)dW_t.\]
If we choose to hold $\varphi(t)$ units of this asset, our portfolio at time
$t$ is worth $\varphi(t)S_t$. The change in the value of our portfolio
in time $dt$ is $d(\varphi(t)S_t)$, and we evaluate this as $\varphi(t)dS_t$,
because we do not change our holding $\varphi(t)$ until after we have seen
the change in the asset price. Here $dS_t=S_{t+dt}-S_t$, so the total
change in the asset over the time-interval $[0,T]$ is the Ito integral
$\int_0^T\varphi(s)dS_s$, in which $\varphi$ is adapted and the stochastic
increment is the forward difference.

We now give a brief account of the Feynman-Kac formula \cite{Kac}.
Feynman related the quantum transition amplitude $\langle\psi,e^{-iHt}\phi
\rangle$ to the integral over histories of $\langle\psi,e^{i\int L(s)\,ds}\phi
\rangle$ where $L$ is the Lagrangian \cite{Feynman}. The trouble is,
the Feynman `integral' over histories is not based on measures, but on
oscillatory integrals, and these rarely converge. In quantum
physics, the spectrum of the energy is bounded below (at least
at zero temperature). This expresses the
stability of the theory. It follows that the unitary time-evolution group
$e^{-iHt}$ has an analytic continuation to complex times with negative
imaginary part. In particular this is true of all the matrix elements
of this operator. This is the underlying fact used in Euclidean
quantum field theory, but also holds for quantum systems without any large
symmetry group; only invariance under time-evolution is needed. In
particular, we can consider the group for negative imaginary times,
giving a semigroup $e^{-Ht}$. The
large-time behaviour of this is very good. This was used by Nelson
\cite{Nelson} to study certain perturbations of the free Hamiltonian: it is
easier to study perturbations of a contraction
semigroup than a unitary group.

\begin{theorem}
Let $H_0=-\frac{1}{2}\frac{\partial ^2}{\partial x^2}$ and $V$ be a
real-valued $C^\infty$-function of $x\in{\bf R}$, vanishing at $\infty$.
Then $H_0+V$ is self-adjoint on Dom$\,H_0$ and
\begin{equation}
\langle\psi,e^{-(H_0+V)t}\varphi\rangle=\int\overline{\psi(\omega(0))}
\varphi(\omega(t))\exp\left(-\int_0^tV(\omega(s))\,ds\right)d\mu.
\end{equation}
\end{theorem}
For the proof, see \cite{Nelson} or \cite{Simon}. For a version within
quantum probability, see \cite{HIK}. In this way, we construct an
interacting theory in terms of a path integral using the Wiener measure
$\mu$, weighted with an exponential function. The similarity with
the Gibbs state of a system of paths in a potential $V$ is noteworthy.
Suppose that $V=0$ outside a region $\Lambda$, and converges to
$+\infty$ inside $\Lambda$. Then we see from Feynman-Kac formula
that the measure vanishes on all paths that enter the region
$\Lambda$. After a normalisation, the weighted measure thus becomes
the conditional Wiener measure, $\mu(\;.\;|\omega(t)\notin\Lambda
\mbox{ for all }t)$. The formula then solves the heat equation
subject to the condition of no-flow through the boundary $\partial
\Lambda$. We do not need to find this conditioned measure to use
the formula; we can, for example, use the Monte Carlo method, and
sample paths by computer, rejecting any that enter $\Lambda$; we
can also use the conditioned measure to get results on monotonicity,
since e. g. if the region $\Lambda$ is enlarged, obviously more
paths are allowed, and so the integral of a positive integrand
is increased. This relation with pde's has developed into the subject
called {\em potential theory} \cite{Grimmett}, and is one of the
tools used in constructive quantum field theory
\cite{Glimm,Glimm2,Frohlich,Guerra}. 

Dyson saw the usefulness of using
imaginary time in quantum field theory \cite{Dyson}.
Schwinger \cite{Schwinger} had introduced
the idea of the Euclidean quantum field as a way
of avoiding the difficulties of Lorentz invariance; these are replaced by
invariance under $O(4)$, the orthogonal group; since we analytically
continue all the time-ordered functions to imaginary time, time $t$ gets
replaced by $it$, often attributed to Minkowski. In fact, Minkowski did not
know
about the consequences of positive energy; he did not analytically continue
anything, but simply replaced time by $-ix_4$, where $x_4=it$. This means
that he considered the complex $O(4)$, and the invariance group was
a particular subgroup $L$ of it
consisting of matrices some of whose entries were complex. In fact, $L$ is
isomorphic to the real Lorentz group, and is thus non-compact. Nothing has
been gained by Minkowski's trick. Indeed, lots of confusion arose in
electromagnetic texts up until recently, where other four-vectors such as
$A^\mu$ were regarded as having a complex zero$^{\rm th}$ component.
Schwinger's programme of Euclidean field theory is a special case of a theory
developed by Wightman \cite{Jost,SW}, in which the expectation values of the
field are
proved to have an analytic continuation in all the space-time components,
into a domain that includes real position variables and purely imaginary time.

Symanzik \cite{Symanzik} started the mathematical programme of Euclidean
quantum field theory. Glimm and Jaffe developed constructive quantum
field theory using their theory of the perturbation of contraction
semigroups. This is almost a Euclidean point of view. A beautiful
probabilistic version of the subject resulted from Nelson's rewrite of
Symanzik's programme. Let us outline this for the quantum mechanics
of an oscillator.

We start with the self-adjoint Hamiltonian
\begin{equation}
H=H_0+V=\frac{1}{2}(-\frac{\partial^2}{\partial q_j^2}+q^2-1)
\end{equation}
Then the lowest eigenvalue, say $0$, is simple; let $U(t)=e^{-iHt}$
and let $\psi_0$ be the eigenfunction of the eigenvalue $0$. Then $\psi_0>0$
holds. That is, there are no nodes in the ground state, a kind of
Perron-Frobenius
theorem. It is then convenient to replace the Hilbert space of the theory,
${\cal H}=L^2({\bf R},dq)$ by the unitarily equivalent space ${\cal
H}^\prime=L^2({\bf R},|\psi_0(q)|^2 dq)$. The unitary map $W:{\cal H}
\rightarrow{\cal H}^\prime$ is given by $(W\psi)(q)=\psi(q)/\psi_0(q)$.
This is obviously organised so that $W\psi_0=1$, the unit constant function
in ${\cal H}^\prime$.
An observable $A$, acting on ${\cal H}$, is converted to $A^\prime=
WAW^{-1}$. The operator $q$ commutes with $W$, so is
unchanged; but its canonical conjugate, $p$ does
not commute with $W$, and neither does $q(t):=U(t)qU(-t)$, so
these operators do not take the usual Schr\"{o}dinger form on ${\cal
H}^\prime$.

The positivity of the energy ensures that the Wightman function
$\langle 1,q(t_1)\ldots q(t_n)1\rangle$
 has an analytic continuation to purely imaginary times, 
\begin{equation}
t_j=is_j,\;\mbox{ such that }s_j-s_{j+1}>0, s_j\in{\bf R},\;j=1,\ldots n-1.
\label{spoints}
\end{equation}
Define the {\em Schwinger function}
\begin{equation}
S_n(s_1,\ldots,s_n)=W_n(is_1,\ldots,is_n)
\end{equation}
at points given by eq.~(\ref{spoints}); we take $S_n$ to be defined
by symmetry in the other regions; since the $w_n$ are symmetric at
real points, the $n!$ analytic functions coincide at a common boundary
of real dimension $n$. So by the edge-of-the-wedge theorem \cite{SW}
there is one common analytic function coinciding with these
Schwinger functions. Obviously, $S_n$ determines $W_n$, by the
uniqueness of analytic continuation.

Then two properties hold: there is a stochastic process $X(t)$
such that $S_n$ is the $n^{\rm th}$ moment:
\[S_n(s_1,\ldots,s_n)=E[X(s_1)\ldots X(s_n)];\]
Moreover, the process is stationary and Markovian; that is
\begin{equation}
E[X_t|{\cal B}_{\leq s}]=E[X_t|{\cal B}_s],\;\mbox{ for }t\geq s.
\end{equation}
Here, ${\cal B}_{\leq s}$ is the $\sigma$-tribe generated by $X_r,\;r\leq
s$, and ${\cal B}_s$ that generated by $X_s$.
Neither of these properties is true for a general Hamiltonian theory, so
they reflect somehow the Lagrangian origins of the theory.

We can recover the physical Hilbert space as the initial space, $L^2(\Omega,
{\cal B}_0,\mu)$ generated by powers $X(0)$ acting on the vacuum, $\psi_0$
which is the function 1. Also $q$ is then multiplication
by $X(0)$. The Hamiltonian can be recovered by the identity
({\em c.f.} (\ref{KBE}))
\begin{equation}
e^{-Ht}P(q)\psi_0=E[P(X(t))|{\cal B}_0]
\end{equation}
for any polynomial $P$.
This is the continuous version of the fact that the transition matrix of
a Markov chain can be recovered as the conditional probability of one
time-step.
We find
\begin{equation}
\langle\psi_0,q(t_1)q(t_2)\psi_0\rangle=(1/2)\exp\{i(t_1-t_2)\}.
\end{equation}
This leads by analytic continuation to
\begin{equation}
S(s_1,s_2)=(1/2)\exp\{-|s_1-s_2|\}=E[X(s_1)X(s_2)]
\end{equation}
where $X(t)$ is the Ornstein-Uhlenbeck process.

Nelson was able to follow this programme for the free quantised field,
and so rewrite the problem of finding solutions to relativistic
quantum fields in terms of generalised random fields. A selection of
good reading on this subject is \cite{Minlos,Wong,Nelson2,Simon2,Gross2}.
\section{Quantum Processes}
Is friction a classical concept? `There is no
friction in quantum systems: the ground state of the atom does not grind
to a halt. The introduction of friction, e. g. the term
$-\gamma\dot{x}$ in Newton's laws, is to account for atomic phenomena
such as radiation of moving charges, in a very crude way. Such effects
are treated exactly in quantum mechanics, and therefore frictional terms
do not appear'. The view is still widespread but not universal among
physicists. Friction does not appear in classical mechanics either if
it is not put in.

A quantum process is, in a general way, a Hilbert space ${\cal H}$
and a family of self-adjoint operators $\{A(t)\}_{t\geq 0}$ on ${\cal H}$.
A quantum field used as {\em noise} appeared in
\cite{Senitzky}. Senitzky obtained
the approximate dynamics of a quantum oscillator by reduction
from the dynamics of a larger conservative system. He arrived at the
following quantum Langevin equation with a Gaussian positive-energy
quantum driving term $(\varphi(t),\pi(t))$ (the noise):
\begin{equation}
\frac{dQ(t)}{dt}=\omega P(t)-\gamma Q(t)+\varphi(t)\hspace{.3in}
\frac{dP(t)}{dt}=-\omega Q(t)-\gamma P(t)+\pi(t).
\end{equation}
He noticed that without the `noise', the Heisenberg commutation
relations fade with time: $[Q(t),P(t)]=ie^{-2\gamma t}$; he considered
this to be inconsistent with quantum mechanics. With the noise, the
solutions obey $[Q(t),P(t)]\approx i$ for all time. The noise
was a free quantum field with constant energy spectrum from $0$ to
$\infty$. This does not quite satisfy the requirement that the Heisenberg
cummutation relations should hold for all time. In \cite{RFS5}
we found the general exact solution to this problem.  A special case is
\[\varphi(t)=2^{-1/2}(a(t)+a^*(t)),\hspace{.4in}\pi(t)=i2^{-1/2}
(a(t)-a^*(t)),\]
where
\[a(t)=(2\gamma/\pi)^{1/2}\int_\omega^\infty e^{-ikt}a(k)\,dk,
\hspace{.3in}[a(k),a^*(k^\prime)]=\delta(k-k^\prime).\]
This has a constant energy spectrum from $\omega$ to $\infty$.
The feature of this solution, and Senitzky's approximate solution,
is the relationship between the dissipation $\gamma$ and the correlation
of the quantum noise, which at zero temperature is 
\[ \langle a(s)a^*(t)\rangle=\frac{2\gamma}{\pi}e^{i\omega(t-s)}\frac{1}{
t-s+i\epsilon}.\]
This is called the fluctuation-dissipation theorem.

Lax \cite{Lax} used noise with all frequencies, with two-point function
\[ \langle a(s)a^*(t)\rangle=\frac{\gamma}{\pi}\delta(t-s).\]
This is closer to the classical white noise, in that the increments to
the process are independent, and the field obeys a quantum version of
the Markov property. It was to be used
later by Hudson and Parthasarathy in a rigorous body of theory
\cite{Hudson3,Partha2}.
As physics, it was criticised by Kubo and others, as violating the
{\em KMS} condition, which comes from the
axiom of positive energy \cite{Haag}. The correct treatment (at
non-zero temperature) was obtained by Ford et al., \cite{Ford} by taking
the limit of one oscillator
coupled to a large system of oscillators (or a string \cite{Lewishb}).
This was truly the quantum Langevin equation, in that the noise is added
only to the equation for $P$ and not to $Q$. This can also be obtained
\cite{HLK} as a singular limit of the asymmetric solution given in
\cite{RFS5}. The quantum noises in \cite{Ford,RFS5} are not martingales,
and have not got independent increments. They do fit in to the axiomatic
scheme offered in \cite{Accardi}. In \cite{Ford2},
Ford emphasizes the role played by causality. Instead of eq.~(\ref{Langevin}),
he considers the equation with memory
\begin{equation}
m\stackrel{..}{x}+\int_{-\infty}^t \mu(t-s)\dot{x}(s)\,ds+V^\prime(x)=F(t).
\end{equation}
The fact that the dissipation due to the future must be zero leads us
to consider only those $\mu$ which vanish for negative argument.
Perhaps this is a lesson for
those \cite{ArakiW,Streater2,RFS3,HIK,Hudson3,Partha2} who like to work
on Lax's version. 

The first work to use the words `continuous
tensor products' ({\em CPT}) was \cite{ArakiW}. The notable conclusion was that the
theory can always be embedded in a boson Fock space; the Wiener chaos is an
example of this. We start with a definition of current algebra, or
better, current group. Let $G$ be a Lie group, with Lie algebra ${\cal G}$,
and denote by ${\cal D}(G)$ the set
of $C^\infty$-maps from ${\bf R}^n$ into $G$, being the identity outside
a compact set. We can furnish ${\cal D}(G)$, the current group,
with a group law by pointwise
multiplication: $fg(x):=f(x)g(x)$. This group has a Lie algebra, denoted
${\cal D}({\cal G})$, which is the set of
all $C^\infty$-maps $F:{\bf R}^n\rightarrow {\cal G}$, of compact support,
under the pointwise bracket
\cite{RFS3}
\[ [F(f),G(g)]:=[F,G](fg).\]
The problem is to find representations of the current groups and
the current algebras, by unitary or self-adjoint operators respectively.

Guichardet \cite{Guichardet} proposed a construction for the
tensor product of Banach spaces or algebras, labelled by a continuous index.
The first thing is to define, if possible, the continuous product of
$f(x)$ over $x\in{\bf R}^n$, when $f$ has compact support. He tries
\begin{equation}
\prod_xf(x):=\exp\left(\int\log f(x)\,dx\right).
\label{prod}
\end{equation}
For Hilbert spaces, we wish to define the scalar product between two
fields of vectors $\psi(x)$ and $\phi(x)$. We put $f(x)=\langle\psi(x),
\phi(x)\rangle$ and use eq.~(\ref{prod}), provided that $f(x)=1$ outside
a compact set and we take $\log 1=0$ (the principal branch).
We then need to be
able to extend the scalar product to linear combinations of product vectors.
In \cite{Dubin}, we
give an example of a non-existent Hilbert continuous product, in that the
positivity fails on linear combinations.
Guichardet presents a class of Hilbert spaces for which the construction
works, and writes the Fock representation of the free field in
these terms. 
To explain his  examples, let ${\cal H}$ be a Hilbert space, and
$\Gamma({\cal H})$ the Fock space over ${\cal H}$. We define the map
$\exp{\cal H}\rightarrow\Gamma({\cal H})$ by
\begin{equation}
\exp\phi:=1\oplus\phi\oplus 2^{-1/2}\phi\otimes\phi\oplus
\ldots\oplus(n!)^{-1/2}\otimes^n\phi\ldots
\label{coherent}
\end{equation}
The $\exp\phi\in\Gamma({\cal H})$ is called the coherent state determined by
the one-particle state $\phi$. One shows that they form a total set (their
span is dense) in $\Gamma({\cal H})$; clearly, 
\begin{equation}
\langle\exp\phi,\exp\psi\rangle=\exp\langle\phi,\psi\rangle.
\label{exp}
\end{equation}
In \cite{Guichardet},
the Hilbert spaces ${\cal H}_x$ at each point is itself the Fock space
$\Gamma(H)$ of a Hilbert space $H$, and the family ${\cal I}$ consists
of coherent states at each point. This is a special case of the construction
given below.

The case of current groups was treated in \cite{RFS2,ArakiW}.
We give here a special case when the continuous label is ${\bf R}$,
interpreted as time; we start with $({\cal H},U,\psi)$, where
${\cal H}$ is a Hilbert space, $\psi\in{\cal H}$, and $U$ is a representation
of $G$ on ${\cal H}$ such $\{U(g)\psi,g\in G\}$ has dense span.
The triple $({\cal H},U,\psi)$ is called a cyclic representation of $G$.

We say that $({\cal H}_1,U_1,\psi_1)$ and $({\cal H}_1,U_2,\psi_2)$
are {\em cyclic equivalent} if there exists a unitary isomorphism
$W:{\cal H}_1\rightarrow{\cal H}_2$ such that for all $g\in G$,
\[WU_1(g)W^{-1}=U_2(g);\hspace{.5in}W\psi_1=\psi_2.\]
A cyclic representation gives us a function on the group, analogous to the
characteristic function of a random variable. Indeed, it reduces to
the characteristic function when the group is ${\bf R}$. Thus
\begin{equation}
C(g):=\langle\psi,U(g)\psi\rangle.
\label{characteristic}
\end{equation}
Let Span$\,G$ denote the complex vector space of finite formal sums of
elements of $G$. Then $C$ is continuous and of positive type on Span$\,G$,
which determines $({\cal H}, U,\psi)$ up to cyclic equivalence.
Conversely, a continuous function $C$ of positive type on $G$
determines a cyclic representation $({\cal H},\pi,\psi)$ related to $C$ by
eq.~(\ref{characteristic}). The construction is very similar to the proof
of the {\em GNS} representation. First, we construct the vector space,
Span$\,G$, and furnish it with the scalar product, determined by its values
on the linearly independent elements $g_1,g_2,\ldots$, by
\[\langle g_i,g_j\rangle=C(g_i^{-1}g_j);\]
we complete Span$\,G$ in the norm (or, if a semi-norm with kernel $K$,
we complete the quotient Span$\,G/K$), giving the space ${\cal H}$.
Then we choose $\psi$ to be
the identity of the group. The operator $U(g)$ can be defined on
Span$\,G$ as left multiplication; this is easily shown to be unitary,
and so can be extended to the whole space to get the representation $U$
of $G$.

In an infinite tensor product over a discrete index, von Neumann was able
to end up with a separable Hilbert space only by labelling a special
vector, say $\psi_x$ in each factor ${\cal H}_x$, and then considering
products
$\otimes\phi_x$ of vectors in a subset $\Delta$ that at infinity are close
to $\psi_x$. Only then
does the infinite product $\prod\langle \phi_x,\psi_x\rangle$ converge.
The tensor product then carries the labels
$\{\psi(x),\Delta\}$. Guichardet used a similar idea for the continuous
product. We are less ambitious, in that
we ask for the tensor product of a cyclic representation $({\cal H},
U,\psi)$ of a group. We use the same representation at each point of
the time axis, because we want to get a stationary quantum
process. We then define the function $C:{\cal D}(G)
\rightarrow{\bf C}$ as
\begin{equation}
C(g(\;.\;):=\prod_x\langle\psi,U(g(x))\psi\rangle,
\label{77}
\end{equation}
which is well defined if we choose at each $x$ one branch of the logarithm.
To get a representation of the current group, it is necessary and sufficient
that this be of positive type on the current group, in which case we say that
the {\em CTP} exists.
We also want the function to be extendable to step functions, constant
in an interval $[s,t]$ and the identity outside. For such a $g(\;.\;)$,
we divide an interval $[s,t]$ into an arbitrary number, $N$, of equal
intervals; then $C(g)$ is the product of $N$
equal factors, each a characteristic function on $G$. Thus $C$
has the property
that it has an $N^{\rm th}$ root that is also a characteristic function.
Such $C$ is called {\em infinitely divisible}. By the relation of
characteristic functions to cyclic representations, we are able to
transfer the concept of $\infty$-divisibility to cyclic representations:
\begin{definition}
Let $({\cal H},U,\psi)$ be a cyclic representation of a group $G$. We
say \cite{RFS2} that it is $\infty$-divisible if, for any integer $N>0$,
there is another cyclic representation $({\cal H}^{1/n},U^{1/n},\psi^{1/n})$,
called the $n^{\rm th}$-root, such
that $({\cal H},U,\psi)$ is cyclically equivalent to
\[\left(\otimes{\cal H}^{1/n},\otimes U^{1/n},\otimes\psi^{1/n}\right)\]
where the tensor product is over $N$ factors, and the resulting
representation is restricted to the cyclic subspace spanned by the group
acting on the product vector $\otimes\psi^{1/n}$.
\end{definition}
We see immediately that if for some $n$ the $n^{\rm th}$ root of the
representation exists, then it is unique (up to cyclic equivalence).
For, the characteristic function of two $n^{\rm th}$-roots, $C_1,C_2$ say,
both satisfy $C_i^n=C$, and so their ratio is $\omega_n$, an
$n^{\rm th}$-root of unity. But this violates positivity unless $\omega_n=1$.
The converse also holds: if $C$ is the product of $n$ functions of positive
type, then $C$ itself is of positive type.
In \cite{RFS2} we assumed that $C(g)$ never vanishes; we prove this later.

Following \cite{RFS2} we can now give the criterion for the positivity
of the scalar product in a continuous tensor product $\otimes^{\psi,\Delta}
{\cal H}_x$ of cyclic group representations,
relative to the cyclic vector $\psi$ and the set of states
$\Delta:=\{U(g)\psi:g\in G\}$.
\begin{theorem}

The following are equivalent.
\begin{enumerate}
\item The function $C(g)$
is a continuous function of positive type on $G$ with
$C(e)=1$, and is $\infty$-divisible.
\item There exists an $\infty$-divisible cyclic representation
$({\cal H},U,\psi)$ of $G$ such that $C(g)=\langle\psi,U(g)\psi\rangle$.
\item $\bigotimes^{\psi,\Delta}$ exists.
\item $C(e)=1$ and a branch of $\log C(g)$ is a conditionally positive
function on $G$.
\end{enumerate}
\end{theorem}
In (3) and (4) the branch of the logarithm is determined by which root
of $C$ is of positive type.
Only the item (4) needs explanation. A function $F(g)$ on a group is said to be
conditionally positive if
\[\sum_{ij}\overline{z}_iz_jF(g_i^{-1}g_j)\geq 0\]
for all $n$-tuples $(g_1,\ldots,g_n)$ of group elements and all
complex $n$-tuples\\
$(z_1,\ldots,z_n)$ summing to zero: $\sum_iz_i=0$.

To sketch the proof, if
$C$ is $\infty$-divisible, and $C=e^F$, then $C^s$ is also of
positive type, for all small $s>0$. Then 
\begin{equation}
\sum_{ij}\overline{z}_iz_j(1+sF_{ij}+\ldots)\geq0,
\label{conditional}
\end{equation}
and so if $\sum_iz_i=0$, we get
that $F$ is conditionally positive semidefinite. For the converse,
if $F$ is conditionally positive definite, then $e^F$ is of
positive type for all $s>0$, see \cite{GelfandV}, page 280.

The following result is called an Araki-Woods
embedding theorem \cite{RFS2}, because of the similarity with
\cite{ArakiW}, (but with different hypotheses).
We remark that under the above
conditions $F$ is conditionally positive semidefinite; then the function
\begin{equation}
\langle g,h\rangle:=F(g^{-1}h)-F(g)-F(h^{-1})
\label{condition2}
\end{equation}
is of positive type, and so can be used to define a
semi-definite form on Span$\,G$ by sesquilinearity.

Let ${\cal K}$ be the (separated, completed) Hilbert space formed using this
as scalar product on Span$\,G$. Let $G_0$ be the subgroup of $G$ such that
$U(g)\psi=e^{i\lambda}\psi$ for some real $\lambda$. We see that
$\langle g,h\rangle$ vanishes on Span$\,G_0$, and
defines a scalar product
on $\mbox{Span}\,G/(\mbox{Span}\,G_0)$, (perhaps after
identifying vectors of zero norm with zero).
We then complete this to give a Hilbert space, ${\cal K}$. We see that the
equivalence class of the identity $e\in G$ is the zero vector
in ${\cal K}$. The original
cyclic representation $({\cal H},U,\psi)$ can then be embedded in the Fock
space over ${\cal K}$, as follows: define the map $W$ from ${\cal H}$ to
$\Gamma({\cal K})$ by its action on the total set $U(G)\psi$:
\begin{equation}
W(U(g)\psi)=C(g)\exp[g],\,g\in G.
\end{equation}
One easily sees that this preserves the scalar product, using
(\ref{condition2}). Thus it can be extended by linearity and continuity
to ${\cal H}$. We see that the cyclic vector $\psi$ is mapped to the
`vacuum' vector $\psi_0$ of the Fock space. As for the group
action, we use the fact that $G/G_0$ is a $g$-space, with left multiplication
$\tau_g[h]=[gh]$. This defines an action $\exp\{\tau_g\}$ on the Fock space
as usual, by its actions on the coherent vectors:
\[\exp\{\tau_g\}\exp[h]:=\exp[gh] .\]
Define an operator $U^\prime$ closely related to $\exp\{\tau_g\}$:
\begin{equation}
U^\prime(g)C(h)\exp[h]:=C(gh)\exp[gh]
\label{embed}
\end{equation}
Then by calculation one sees that $({\cal H},U,\psi)$ is cyclically
equivalent to the cyclic subspace of $({\cal K},U^\prime,\psi_0)$; $W$
intertwines $U$ and $U^\prime$ and maps $\psi$ to $\psi_0=\exp[e]$.
From the unitarity of $U^\prime$ we see that
$|C(g)|^2=e^{-\langle[g],[g]\rangle}\neq0$.

The Gaussian measure is $\infty$-divisible, and the representation
of the translation group, $U(\lambda)$,
with Gaussian cyclic vector $\psi(x)=(2\pi)^{-1/4}e^{-x^2/4}$, is
$\infty$-divisible. The corresponding {\em CTP} contains Brownian
motion  \S2; the continuous product $\otimes_0^tU(\lambda)$ is the
exponential martingale. A representation
of the oscillator group is $\infty$-divisible, and the {\em CTP}
of this is the free non-relativistic quantised fields \cite{RFS3}.

H. Araki independently obtained similar results \cite{Araki}. Instead of
$\infty$-divisible cyclic representations of groups, Araki started with
a factorizable representation of current algebra.
He remarked that, putting $[g]=\phi_g$
the map $V(g)\phi_h:=\phi_{gh}-\phi_g$ 
is a unitary representation of $G$; this is proved on the vectors
$\phi_h$, $\phi_k$ by use of (\ref{condition2}). The equation
expresses that the
map $g\mapsto \phi_g\in{\cal K}$ is a {\em one-cocycle} of the group, with
values in ${\cal K}$. We briefly explain this.

So, let $G$ be a group, and let ${\cal K}$ be a Hilbert space on which
$G$ acts by unitary operators $g\mapsto V(g)$.
We shall write the left action $\phi\mapsto V(g)\phi$ as left
multiplication, $\phi\mapsto g\phi$. The right action, which appears in the
general theory of group cohomology, is taken to be trivial: $\phi g=\phi$.
An $n$-{\em cochain} with values in
${\cal K}$ is a map from
$G^n$ into ${\cal K}$, that is, it is a function of $n$ group elements
with values in ${\cal K}$, thus: $\phi(g_1,\ldots,g_n)$. We shall need
only the $0$-cochains, which make up the space $C^0:={\cal K}$ of vectors
independent of $g$, and the 1-cochains, which are vector fields $\phi(g)
\in{\cal K}$ defined on the group. These make up the vector space $C^1$.
We shall also need the  $2$-co-chains, when
${\cal K}={\bf C}$; these are complex-valued functions of two group
elements. We see that the cochains of any degree $k$ form a vector space
$C^k$. Fundamental to any cohomology theory is the coboundary operator,
which is a linear map, $\delta:C^k\rightarrow C^{k+1}$,
so increasing the degree of the cochain. It obeys $\delta^2=0$.
In the case of a group $G$ and a left and right action of $G$ on ${\cal K}$,
$\delta$ is the linear map defined on $C^0$ by
\[(\delta\phi_0)(g)=g\phi_0-\phi_0g.\]
On $C^1$, $\delta$ is the linear map defined by
\[(\delta\phi_1)(g_1,g_2)=g_1\phi_1(g_2)-\phi_1(g_1g_2)_+\phi_1(g_1)g_2.\]
On $C^2$, $\delta$ is the linear map defined by
\[(\delta\phi_2)(g_1,g_2,g_3)=g_1\phi_2(g_2,g_3)-\phi_2(g_1g_2,g_3)+
\phi_2(g_1,g_2g_3)-\phi_2(g_1,g_2)g_3.\]
The vector space of cocycles of degree $k$ in a vector space ${\cal K}$, with
left and right actions $\tau_1,\tau_2$, is denoted $Z^k(G,{\cal K},\tau_1,
\tau_2)$. One checks that $\delta^2=0$. A coboundary of degree $k$
is a vector function
of the form $\delta\psi$, where $\phi$ is a cochain of degree $k-1$.
The coboundaries of degree $k$ form the vector space $B^k(G,V,
\tau_1,\tau_2)$. Since $\delta^2=0$, we see that every coboundary is a
cocycle. If the converse holds, the cohomology group $H^k:=Z^k/B^k$,
is trivial.
One sees that if $\phi$ is a one-cocycle in $C^1(G,
{\cal K},V)$, then $\langle\phi(g_1^{-1}),\phi(g_2)\rangle$ is a two-cocycle
in $C^2(G,{\bf C},I)$.

A $2$-cocycle $\sigma(g,h)$ with values in the unit circle is also
called a multiplier for the group. A multiplier representation of a group
$G$ is a map $g\mapsto U(g),\;g\in G$, such that $U(g)U(h)=\sigma(g,h)U(gh)$
for all $g,h\in G$. Although Wigner's analysis of symmetry in quantum
mechanics leads naturally to multiplier representations, their occurrence
is sometimes called an `anomaly' by physicists.
When the {\em CTP} exists, we can represent the element
$g(\;.\;)$ of the current group by the operator $(\otimes U)_g$, defined on
the product vectors $\otimes_x U(h(x))\psi_x$ by
\begin{equation}
(\otimes U)_g(\otimes_x U(h(x))\psi_x):=\otimes_xU(g(x)h(x))\psi_x,
\label{current}
\end{equation}
The space of the {\em CTP} is then $\Gamma(\int_{\oplus}\exp{\cal K}dx)$
and $\Delta$ consists of coherent states of the form $\exp\phi_{g(x)}$.
So we obtain a local representation of the current algebra. We get a
multiplier when the branch of the logarithm in (\ref{77}) obtained by the
group law differs from the one needed to give a function of positive type on
the group. This gives rise to an anomaly.

Araki showed that if $\phi$ is the cocycle defined by the $\infty$-divisible
representation $U$, then it is necessary that ${\rm Im}\langle
\phi(g_1^{-1}),\phi(g_2)\rangle$ be a coboundary. Conversely, given a
cocycle $\phi$
with this property, it comes from an $\infty$-divisible representation.
He proved that if $G$ is compact, then any cocycle is a coboundary, i. e.
of the form $\phi_g=(V(g)-I)\chi$ for some $\chi\in{\cal K}$. Use of
a coboundary leads to a {\em CTP} of the form assumed by
Guichardet \cite{Guichardet}.
Araki was able to obtain
analogues of the Levy formula (\ref{Levy}) for various groups;
for the group ${\bf R}$ this takes on
a new meaning, as the decomposition of a cocycle into its parts
coming from primitive cocycles, some algebraic and some topological.
The topological cocycles are of the form $(V(g)-I)\chi$; it is not a
coboundary because $\chi$ is not in ${\cal K}$, but lies in a larger
space that admits an extension of $V$; the $V(g)-I$ brings the
vector back into ${\cal K}$. Some groups, e. g. ${\bf R}$, also have
cocycles called algebraic by Araki. For example, in the case $G={\bf R}$,
take ${\cal K}={\bf C}$, and $V(a)=I$ for all $a\in G$. The cocycle is
$\phi(a)=a$. Then $\langle\phi_a,\phi_b\rangle=ab$ is real, and $C(\lambda)
=\exp\{-\frac{1}{2}\lambda^2\}$, the characteristic function of the Gaussian
distribution. The Poisson part of the Levy formula comes from
the coboundaries, and the Levy processes from the topological cocycles.

The question arises, given ${\cal K},V$ and a cocycle $g\mapsto \phi_g$,
can we construct a {\em CTP}? We can construct $({\cal H},
U(g),\psi)$ from $C$, which can be regarded as a function such that
$C(e)=1$ and the map $C(h)\exp\phi_h\mapsto C(gh)\exp\phi_{gh}$ is unitary.
The next big step was by
Parthasarathy and Schmidt \cite{Partha}, who showed that given a cocycle
there is indeed an $\infty$-divisible representation associated with it,
but that it is a multiplier representation, with an $\infty$-divisible
multiplier $\sigma$. The corresponding function $C(g)$ is $\sigma$-positive.
This means that
\begin{equation}
\sum_{ij}\overline{z}_iz_j\sigma(g_i^{-1},g_j)C(g_i^{-1}g_j)\geq0.
\end{equation}
Naturally, this gives to a multiplier representation of the current group
in general, and they found the multiplier; this leads to a tidier theory than
\cite{Araki}, since the condition for the absence of multiplier can be
dropped. Since the physical interpretation of a symmetry group
leads (according to Wigner\cite{Wigner}) to the ambiguity of the
induced unitary representation up to a coboundary, the projective theory
is certainly the right setting. Holevo has presented some similar
concepts at the level of the algebra of observables, and found applications
in quantum theory \cite{Holevo}. Notable in the development was the work of Gelfand, et al.
\cite{Gelfand} who used a cocycle of $SL(2,{\bf R})$ to construct a
factorisable representation of the corresponding current group.
The whole theory is well explained in \cite{Guichardet2,Erven}.

A theory of processes with independent increments
and values in a Lie algebra ${\cal G}$ was developed
in \cite{RFS4}, extended to multiplier representations by Mathon
\cite{Mathon} and to Clifford algebras in \cite{Mathon2}.
Corresponding central limit theorems were proved by Hudson, and Cushen
and Hudson \cite{Cushen,Hudson7}.
A Lie process
can be obtained by differentiation of the corresponding object for a
Lie group. For example, near the identity any group element $g$ lies on a
one-parameter subgroup generated by an $X\in{\cal G}$, and we
write (Exp means the exponential map from ${\cal G}$ to $G$, not
the Fock map)
$g(t)=\mbox{Exp}\,tX,\;g(0)=e,\;g(1)=g$; given a representation $U(g)$ we get
a representation of ${\cal G}$ by $\pi(X)=d/dt[U(g(t)]_{t=0}$. By Stone's
theorem, $X$ is self-sdjoint. However, given a cyclic vector $\psi$  for $U$
it does not follow that $\psi$ is cyclic for $\pi$, because of domain
questions. Let ${\cal E}$ be the universal enveloping algebra of ${\cal G}$.
This is the nonabelian polynomial algebra, modulo the ideal generated by
the commutators $XY-YX-[X,Y]$. Here, $[X,Y]\in{\cal G}$ is the Lie product,
a polynomial of degree 1. A cyclic representation $({\cal H},\pi,\psi)$
is determined
(up to equivalence) by a positive linear functional, or state, on ${\cal E}$:
\[X_1X_2\ldots X_n\mapsto \langle\psi,\pi(X_1)\pi(X_2)\ldots\pi(X_n)\psi\rangle
=W_n(X_1\ldots X_n).\]
These are the noncommutative moments, or Wightman functions; they determine
a representation, by the Wightman reconstruction theorem \cite{SW}.
They are generated by the characteristic function
\begin{equation}
C({\lambda})=\langle\psi,U(\mbox{Exp}\lambda_1X_1)\ldots U(\mbox{Exp}
\lambda_nX_n)\psi\rangle,\;\;\lambda\in{\bf R}^n.
\label{Hegerfeldt}
\end{equation}
Here, $\{X_j\}$ is a basis of the Lie algebra, and any moment out of order
is determined by a derivative of $C$ and use of the commutation relations.
The truncated functions $W_T$ are generated by $\log C$, \cite{Hegerfeldt2}
and are related to
$W$ by a formula similar to eq.~(\ref{cumulants}), relating cumulants to the
moments. Two cyclic representations with the same $W$, or the same $W_T$,
are cyclic equivalent. The cumulants of $\exp U$ (the Fock construction)
are the same as the moments of $U$; this follows from $\exp U(g)\exp U(h)
\psi)=\exp(U(gh)\psi)$ and (\ref{Hegerfeldt}). 

Given two representations $U_1$, $U_2$ of $G$, their tensor product
$U_1\otimes U_2$, restricted to the diagonal subgroup of $G\times G$, gives
the representation $\pi_1\otimes I+I\otimes\pi_2$ of ${\cal G}$. This led to
the use of a coproduct, though it was not recognised as such until
\cite{Schurmann}. Whereas a product on an algebra ${\cal A}$ is a linear
map ${\cal A}\otimes{\cal A}\rightarrow{\cal A}$, a coproduct is a map
${\cal A}\rightarrow{\cal A}\otimes{\cal A}$.
For Lie algebras the coproduct is $X\mapsto X\otimes I
+I\otimes X$. Then we say that a cyclic representation $({\cal H},\pi,\psi)$
is $\infty$-divisible if for each $N$ there is another, $({\cal H}^{1/N},
\pi^{1/n},\psi^{1/N})$ such that $({\cal H},\pi,\psi)$ is cyclically equivalent
to $\pi^{1/N}\otimes I+I\otimes\pi^{1-1/N}$. Starting at $N=2$ this gives
the concept of rational powers of $\pi$.

The differentiation of a {\em CTP} representation $\otimes_t
U_t(g(t))$ of the
current group leads to an {\em ultralocal field} \cite{Arakithesis,Klauder}. These are
such that the truncated Wightman functions have the form
\begin{equation}
W_T(X_1(f_1)\ldots X_n(f_n))=\kappa_n(X_1\ldots X_n)\int f_1(t)\ldots
f_n(t)\,dt.
\label{ultralocal}
\end{equation}
Here, $\{\kappa_n\}$ are the cumulants of $\pi=dU$. The commutative analogue
was analysed in \cite{GelfandV}. For Lie algebras, we found \cite{RFS4}:
\begin{theorem}
The following are equivalent;\\
1) Eq.~(\ref{ultralocal}) defines a representation of ${\cal D}({\cal G})$.\\
2) The $\kappa_n$ are the cumulants of some $\infty$-divisible cyclic
representation of ${\cal G}$.\\
3) The $\kappa_n$ are positive semi-definite on ${\cal E}_1$,
the subalgebra of ${\cal E}$ with identity omitted.
\end{theorem}
We note that (3) is the expression of conditional positivity at the
algebraic level.
Since the cumulants of $\exp U$ are the moments of $U$, we can get a set of
$\kappa_n$ that obey the positivity (3) by using the moments of $\exp U$.
These happen to have a positive extension to ${\cal E}$: any
conditionally positive functional is positive. Th. (5.3) has a cohomological
version, which we outline.

Let ${\cal E}$ be an associative algebra with identity, ${\cal K}$ a linear
space and
$\tau$ a representation of ${\cal E}$ on ${\cal K}$. The $p$-cochain group
$C^p({\cal E},{\cal K},\tau)$ is the linear space of $p$-multilinear maps
$\phi:{\cal E}\times\ldots{\cal E}\rightarrow{\cal K}$. The coboundary
operator $\delta:C^p\rightarrow C^{p+1}$ is given by
\[(\delta\phi)(X_1,\ldots,X_{p+1})=\tau(X_1)\phi(X_2,\ldots,X_{p+1})+
\sum(-1)^j\phi(X_1,\ldots,X_jX_{j+1},\ldots,X_{p+1}).\]
Then $\delta^2=0$ and we define as usual the cocycle group $Z^*:=\mbox{ker}\,
\delta$ and the coboundary group $B^*:=\mbox{Ran}\,\delta$, and the
cohomology as $H^*:=Z^*/B^*$. ($^*$ means for any $p$). We see that
a 1-cocycle is a map $\phi:{\cal E}\rightarrow{\cal K}$ that satisfies
$\phi(XY)=\tau(X)\phi(Y)$, and a 1-coboundary 
is a cocycle of the form $\phi(X)=\tau(X)\phi_0$ for some $\phi_0\in
{\cal K}$.

The states on ${\cal E}_1$ are positive elements of $B^2
({\cal E}_1,{\bf C},0)$. Thus if $({\cal H},\pi,\psi)$ is
$\infty$-divisible, then its cumulants $W_T$ define a state on ${\cal E}_1$, and thus
a scalar product:
$\langle X,Y\rangle:=W_T(X^*Y)$. Here we define $X^*=-X$, since we want $\pi$
to represent the generators $iX$ of one-parameter subgroups by hermitian
operators. Define ${\cal K}$ as the separated prehilbert space obtained from
${\cal E}_1$ as usual. Let $\phi:{\cal E}_1\rightarrow{\cal K}$ be the
embedding obtained from this, and define a *-action $\tau$ of ${\cal G}$
on monomials by
\[\tau(X)\phi(X_1\ldots X_n):=\phi(XX_1\ldots X_n).\]
This states that $\phi$ is a 1-cocycle. We then show that there is a
bijection between the set of $\infty$-divisible cyclic representations
$({\cal H},\pi,\psi)$ of ${\cal G}$ and the triples $(\tau,\phi,\chi)$,
where $\tau$ is a hermitian representation of ${\cal G}$ on a prehilbert
space ${\cal K}$, $\chi$ is a real character, and
$\phi\in Z^1({\cal E}_1,{\cal K},\tau)$ such that
\begin{equation}
\gamma:=\mbox{Im}\langle\phi(X),\phi(Y)\rangle\in B^2({\cal E}_1,{\bf R},0).
\label{araki}
\end{equation}
In this bijection, ${\cal H}$ is embedded in $\Gamma({\cal K)}$, $\psi$ is
mapped to the Fock vacuum, and $\pi$ is related to $\exp\tau$ \cite{RFS4}.
So this is the Araki-Woods embedding theorem in this case.
If (\ref{araki}) fails then we get a projective representation of ${\cal G}$,
with multiplier $\sigma$ related to the cocycle $\gamma$
\cite{Mathon,Erven}. We see that a cocycle for ${\bf R}$ is defined by
a function $\chi\in L^1({\bf R})$ such that $x\chi\in L^2({\bf R})$.
We thus see the origin of the condition near $\alpha=0$ in (\ref{Levy}).

In \cite{Mathon2} we show that for Clifford algebras, the only possible
$\infty$-divisible states are
Gaussian (all cumulants above the second vanish). Here 
the coproduct is that of Chevalley,
$A\mapsto A\otimes I+(-1)^FI\otimes A$ where $F$ is the
degree of $A$, for elements of even or odd degree.

The algebraic theory
was extended to associative algebras (that were not enveloping
algebras of Lie algebras) by Hegerfeldt, who applied it to classify
$\infty$-divisible quantum fields \cite{Hegerfeldt2}.

Goldin et al. have, independently of this work, constructed
representations of a vector form of charge-current algebra, starting
with the Fock space creation-annihilation operators \cite{Goldin}; they
have been able to identify the representations in terms of the general
anlysis of semi-direct products.

Sch\"{u}rmann \cite{Schurmann} introduced the concept of infinite
divisibility for a representation of a Hopf algebra, and obtained 
essentially all the results of \cite{RFS4,Mathon,Mathon2} in this more
general setting.
Stochastic integrals for these processes were also constructed.
For a clear account, see \cite{Meyer}.

Voiculecsu developed the algebraic side into a subject called `free
probability' \cite{Voiculescu}, as it lives in full Fock space, without
symmetry or antsymmetry.

Albeverio and Hoegh-Krohn \cite{Albeverio} have constructed
representations of current 
groups, and been able to replace the independence at every point by
a covariance similar to the Nelson free field.
\section{Quantum Stochastic Semigroups}
These models of non-commutative noise, or quantum noise, are possible
driving random terms for noisy quantum dynamics.
What should we be looking for in a nonequilibrium stochastic quantum
dynamics? From 1970, E. B. Davies made progress in formulating stochastic
quantum dynamics \cite{Davies}.
Suppose that the $C^*$-algebra of observables is ${\cal A}$.
We look at the Fokker-Planck equation in the classical case, and we see that
we might expect a quantum stochastic process to be determined by a
semigroup (in continuous or discrete time) of maps $T_t$
from the state space $\Sigma({\cal A})$ to itself. It must map positive
operators, the density matrices, to positive operators, and preserve the
trace. We also do not want it to map a normal state to one of the
finitely additive ones, so we require a {\em stochastic map} to obey
\begin{enumerate}
\item $T$ maps $\Sigma$ to itself;
\item $T$ is linear;
\item In continuous time, $\|(T_t-I)A\|_1\rightarrow 0$ as $t\rightarrow0$.
\end{enumerate}
We can throw the action onto to algebra, to get the dual action $T^*:
{\cal A}\rightarrow{\cal A}$, by the requirement that for $A\in{\cal A}$,
\[\langle T\rho,A\rangle=\langle\rho,T^*A\rangle\mbox{ for all }\rho\in\Sigma.\]
$T^*$ is automatically normal. We see that if ${\cal A}$ is abelian,
then our conditions reduce to the properties
needed for a classical stochastic process. It is obvious that a
unitary time-evolution gives us a one-parameter family of
stochastic maps, which can be extended to a group by including the
inverses. We can get a large class of stochastic maps by
forming mixtures of unitary groups; thus if $\tau_j$ is a family of
invertible dynamics, then $T=\sum_j\lambda_j\tau_j$ is stochastic if
$\lambda_j\geq0$ and $\sum\lambda_j=1$.
Any stochastic map is non-invertible if
it is not unitary, and so is in this sense dissipative \cite{Davies}, p 25.
In addition, in the quantum case, Kraus \cite{Kraus} has argued that
to get a satisfactory interpretation of the semigroup, $T$ must be
{\em completely positive}. We say that a map $T:{\cal A}\mapsto{\cal A}$
is $n$-positive if $T\otimes I_n$ is positive on the algebra ${\cal A}\otimes{\bf
M}^n$. This is needed, since if our quantum system is described by the
algebra ${\cal A}$, and there is an $n$-state quantum system far away,
then the combined system will be described by ${\cal A}\otimes{\bf M}^n$,
and the dynamics on the combined system could be $T\otimes I_n$. This
must be positivity preserving, or else some state of the combined system
will evolve to give negative probabilities. Since we want to avoid this for
all $n$, we want $T$ to be $n$-positive for all $n=1,2\ldots$. Such a
condition is called complete positivity. It should be said that any
positive map on an abelian algebra is always completely positive, so this
concept only seriously arises in quantum probability.

Kraus showed that a map $T$ is completely positive if and only if $T(A)$
is a sum of maps of the form $S_n^*AS_n$, where the $S_n$ are bounded;
\cite{Davies}, p. 140.

The great result in the subject is the classification of continuous
semi-groups of completely positive maps. In finite dimensions this was
achieved in \cite{Gorini}, and independently, by Lindblad, \cite{Lindblad}
whose
result holds for norm-continuous dynamics on $C^*$ algebras. Their result
is the quantum analogue of the heat equation, i. e. it is a dynamical
equation for the density matrix. For a simple derivation,
see \cite{Landau2}. The result is:
\begin{theorem}
Let $T_t$ be a semigroup of completely positive stochastic maps on ${\bf
M}^n$. Then there exists a Hermitian matrix $H$ and matrices $S_j$
such that the generator of the semigroup has the form
\begin{equation}
Z(A)=i[H,A]-\frac{1}{2}(RA+AR)+\sum_jS_j^*AS_j,\hspace{.3in}\mbox{ where }R=
\sum_jS^*_jS_j.
\label{Lindblad}
\end{equation}
\end{theorem}
This can be thrown onto the density matrices by duality. The first term
$i[H,A]$ is non-dissipative, and is called the hamiltonian term.
The second term is the dissipation.

It is very interesting that the first two terms
of the Heisenberg expansion of the dynamics are of this form. Thus,
\begin{eqnarray*}
\left(e^{iHt}Ae^{-iHt}-A\right)t^{-1}&=&i[H,A]-\frac{1}{2}[H,[H,A]]t+
O(t^2)\\
&=&i[H,A]-\frac{1}{2}(AS^2+S^2A)+SAS\\
& &\mbox{ where }S=Ht^{1/2},
\end{eqnarray*}
up to O(t), so it is of the form eq.~(\ref{Lindblad}) with $R=S^2$.
In the {\em anti-van Hove limit} \cite{Streater} we replace
$S$ by $\lambda H$.

It has been remarked that the commutator $A\mapsto i[H,A]$ is a derivation
of the operator algebra, and so has many of the properties of a derivative.
The double commutator
has many of the properties of the second derivative, including some
positivity, which mimics the positive spectrum of $-\Delta$ and the
positivity improving properties of $e^{\Delta t}$.
Lindblad has analysed continuous semigroups of cp maps, with generator ${\cal
L}$, in terms of the `dissipation operator', being minus the coboundary of
$L$:
\begin{equation}
D(A,B):=-\delta{\cal L}(A,B)={\cal L}(AB)-{\cal L}(A)B-A{\cal L}(B).
\end{equation}
He proves that $T_t:=\exp(i{\cal L}t)$ is a continuous semigroup of cp maps
if and only if $D$ is positive in the sense that
\begin{equation}
\sum_{ij}C_i^*D(A_i^*,A_j)C_j\geq0\hspace{.3in}\mbox{ for all }A_i, C_j\in
{\cal A}.
\end{equation}
Note the formal similarity with \cite{Araki,RFS2,RFS4,Partha}. Fannes and
Quaegebeur \cite{Fannes} have defined the concept of $\infty$-divisible
completely positive mappings on groups, in which the function $C(g)$ is
replaced by a cp operator. They prove an Araki-Woods embedding theorem
for such structures.

Recall that for Markov chains, Brownian motion and Euclidean field theory,
we can express the given semigroup as an isometric time-translation,
followed by the conditional expectation onto the initial space.
By using two-sided time, the isometries can be replaced by a unitary group.
The finding of the appropriate unitary group is called the {\em dilation}
of the semi-group. It is not unique, but there is a unique minimal one.
\cite{Stinespring}. It would be nice to interpret the dilated system
as representing the full physics of system plus environment, with a
unitary evolution; the projection onto a subspace represents our loss
of information due to incomplete knowledge. The ambiguity of the dilation
then shows that several different models give the same (crude)
coarse-grained dynamics. However, it will rarely be the case that a dilation
has the good properties, such as positivity of the energy, needed for
this interpretation.

This is illustrated in the quantum case, which in finite dimensions
takes the form \cite{Davies}
\begin{theorem}
Let $T_t$ be a semigroup of cp stochastic maps on ${\bf M}_n$ acting
on ${\cal H}$. Then there exists a Hilbert space ${\cal K}$, a pure state
$\rho$ on ${\cal H}\otimes{\cal K}$ and a one parameter unitary group
$V_t$ on ${\cal H}\otimes{\cal K}$ such that
\[ T_t(A)=E_\rho[V^*_t(A\otimes I)V_t]\]
for all $A\in{\cal M}$ and all $t\in{\bf R}$.
\end{theorem}

This is proved by putting together Theorem (4.2) and \S 7.2 of \cite{Davies}.
Note that the Hilbert space ${\cal K}$ is constructed by adding Wiener
noise, and so is not finite-dimensional. The semi-group has been
dilated to a unitary group on the Wiener space with two-sided time;
the generator of time-evolution is not bounded below, since it has white
spectrum. This does not represent an environment at any finite temperature.
A special case is the dilation of the semigroup given by the anti-van Hove
limit. In that case the process is given by
\begin{equation}
X(t)=(2\pi t\lambda^2)^{-1/2}\int e^{-s^2/(2\lambda^2t)}U(s+t)XU(-s-t)\,ds.
\end{equation}
This has the interpretation as the Heisenberg evolution, but with the time
$t$ slightly uncertain, and getting more uncertain in the future.
This interpretation is only a slight
variation on the methods used in the justification of the microcanonical
state by ergodic theory. There, it is said that the atomic times are so
small that we never measure an observable {\em at} a particular time;
rather, we measure the average over the time $0\leq s\leq t$
of the measurement thus: $\overline{A}=t^{-1}\int_0^t A(s)ds$. Since
$t$ is so large compared with the atomic processes, we take the limit
$t\rightarrow\infty$. This idea is a non-starter for non-equilibrium
statistical mechanics, since if the limit exists it is time-independent.
Instead, we may say that we cannot measure an observable at an {\em exact}
time, but form the weighted average, with Gaussian weight, around
the desired time $t$. The uncertainty in the Gaussian is $\lambda^2 t$,
growing with time. $\lambda$ is the dissipation parameter. In models it
turns out to be the hopping parameter of the atomic system.

Some authors limit the concept of quantum stochastic process to
the case where the possible observed path of measurements themselves
make up a classical process. The grounds for this is that the observations
(in a set of repeated experiments) have actually been seen; these form
the {\em quantum record}; take them to
form a sample space. However, this is not true. The process $X(t)$
at different times might not
commute, so the measurement of $X(t)$ alters the state (by collapse),
and subsequent measurements are not those predicted by $X(t+s)$, $s>0$,
as computed using the given initial state. It needs conditioning
to the new information, and quantum conditional expectations only
commute on abelian subalgebras.
Moreover, one can measure $X(t)$ in one sampling and $Y(t)$ in another,
where $X$ and $Y$ do not commute. No classical model would predict
the statistics of the process; the classical theorist is liable to be hit
by the {\em EPR} paradox in acute form. We regard $X(t)$ as the observable
seen at time $t$ when no measurement has been made in $\{s:0<s<t\}$. So we
cannot agree with the idea that the randomness itself is caused by the
reduction of the wave-function due to continuous measurement; it might
be due to interaction with a large other body, but not one designed to
measure any particular observable.

Davies's dilation of the Lindblad semigroup uses a number
of independent Wiener processes to provide the set-up.
The question arises whether there is a relation between quantum dynamical
semigroups and a class of quantum stochastic differential equations, similar
to the relation between the Fokker-Planck equation (\ref{63}) and the sde
(\ref{sde}). For this, we need a quantum version of Ito's integral.
In 1956, Umegaki defined the concept of conditional expectation in
non-commutative integration theory \cite{Umegaki}. Let ${\cal A}$ be a
von Neumann algebra
with a semi-finite trace, and say an operator $A$ is integrable if
$\mbox{Tr}\,|A|<\infty$. The vector space of integrable operators can
be completed to form the space $L^1({\cal A})$. Segal and Nelson showed
that there
is a closed operator representing an element of the completion.
Let ${\cal A}_t$ be an increasing family
of subalgebras which generate ${\cal A}$ and are right continuous
\cite{Barnett}, such that the trace, restricted to each ${\cal A}_t$
is semi-finite.
Then a conditional expectation relative to the trace
is a linear map $M:L^1({\cal A})\rightarrow L^1({\cal A}_t),\;t\geq 0$,
such that
\[\mbox{Tr}(XA)=\mbox{Tr}(M_t(X)A)\hspace{.5in}\mbox{ for all }A\in{\cal
A}_t,\hspace{.1in}X\in L^1({\cal A}).\]
A martingale is a process $X_t$ of integrable operators such that
\[M_sX_t=X_s\]
for all $0\leq s\leq t$.
This concept can be generalised to a filtration of algebra with specified
state, rather than trace.

Cuculescu \cite{Cuculescu} proved a martingale convergence theorem for
discrete time. Barnett \cite{Barnett} obtained a martingale theorem for
continuous time. This work persuaded us to look for examples of noncommuting
martingales. Soon we found plenty within the theory of continuous
tensor products \cite{Hudson}. Let $({\cal H},U,\psi)$ be an
$\infty$-divisible representation of a Lie group G, and consider
$\otimes_{t=0}^\infty {\cal H}_t$ relative to the vector $\otimes\psi_t$
and the set $\Delta$ of coherent vectors. Here, all factors are the same.
To $g\in G$ we associate the family of unitary operators
\begin{equation}
V_t(g):=\otimes_0^t U(g)\otimes_t^\infty I.
\label{mart}
\end{equation}
We call such an operator {\em simple}, localised in $[0,t]$.
Let ${\cal A}_t$ be the algebra generated by $\{V_s(g)\}$ with
$0\leq s\leq t$ and $g\in G$. Then for $s<t$ define the map $M_s:{\cal A}_t
\rightarrow{\cal A}_s$ by continuous linear extension of $M_s\otimes_{r=0}
^tV_r(g)=\otimes_{r=0}^sV_r(g)$. Then $M_s$ is a conditional expectation,
and relative to $M_s$, the family $V_t$ is a martingale.
Applied to $G={\bf R}$ with $\psi$ a Gaussian state, $V_t$ is the
exponential martingale of Brownian motion.
When $G$ is the oscillator group, the lie algebra is spanned by $P,Q,H$ and
a central element $I$. There is a representation by self-adjoint operators
on $L^2({\bf R})$, with the ground state of the harmonic oscillator as cyclic
vector. This is
infinitely divisible, and the unitary operators in (\ref{mart})
are copies of the exponential martingale $e^{iW_t}$ for the subgroups
generated by $P$ and $Q$, and is the Poisson exponential martingale
for the subgroup generated by $H$ \cite{Wulfsohn}. This became known as the
gauge process \cite{Partha2}.
All these martingales are defined on the total set of coherent states.
Since they are unitary, they can be extended to an everywhere-defined
unitary group, the generators of which are self-adjoint operators.
This is
the main technique of the Hudson-Parthasarathy calculus
\cite{Hudson4,Hudson3,Partha2}

Examples of martingales with trace were given in \cite{Barnett2}.
Consider the Fock Fermi operators $b(f),b*(g)$ with anticommutation relations
$[b(f),b^*(g)]=\langle f,g\rangle$ for $f,g\in L^2({\bf R}_+)$. The algebra
generated by these and the Fock condition $b(f)|0\rangle=0$ is represented on
antisymmetric Fock space over $L^2({\bf R}_+)$ as the $W^*$-algebra
generated by the Fermi field $\psi(f)=b(f)+b^*(\overline{f})$ acting on the
Fock vacuum $|0\rangle$. The Clifford process is the set of operators
\begin{equation}
\Psi(t):=\psi(\xi_{[0,t]}).
\end{equation}
The non-commutative integration theory \cite{Segal0,Segal2,Kunze},
taking the place of measure theory, is that based on the hyperfinite
von Neumann factor of type $II_1$, which is
furnished with a faithful trace $\varphi(A)=\langle 0|A|0\rangle$.
The completion of ${\cal A}$ in the norm $\|A\|=\varphi(A^*A)^{1/2}$
is denoted $L^2({\cal A},\varphi)$.
The projection $M_t$ from $L^2({\cal A},\varphi)$ onto $L^2({\cal A}_t,
\varphi)$
is the same as the projection from $\Gamma(L^2[0,\infty])$ onto $\Gamma
(L^2[0,t])$; it obeys the laws for a conditional expectation, and
$\Psi(t)$ is a martingale.

The increments of $\Psi(t)$ are independent, but anti-commute. Otherwise, all
the properties are analogous to Brownian motion. The isometric time-evolution
analogous to the left shift of the classical theory is that given by the map
$U_s:\Psi(t)\mapsto \Psi(s+t)$. The antisymmetric Fock space over
$L^2({\bf R})$ carries a unitary extension of $U_s$, namely the second
quantisation of translation in ${\bf R}$.
We define an {\em adapted} process $h(t)$ to be a family of operators
such that $h(t)\in{\cal A}_t$; it is {\em simple} if it can be expressed
as
\begin{equation}
h=\sum_{k=1}^n h_{k-1}\chi_{[t_{k-1},t_k)}\mbox{ on }[0,t).
\end{equation}
We then define the stochastic integral of any simple adapted process,
relative to $\Psi$, to be that constructed in the manner of Ito,
with the forward difference in $d\Psi$:
\begin{equation}
\int_0^t f(s)d\Psi(s):=\sum_{k=1}^n h_{k-1}\left(\Psi(t_k)-\Psi(t_{k-1})
\right).
\label{Itoclifford}
\end{equation}
As in Ito's theory, what make it work is an isometry property:
\begin{theorem}
If $h(t)$ is a simple process made up of $L^2$ operators, then $\int_0^t
h(s)d\Psi(s)\in L^2$, and
\[\|\int_0^th(s)d\Psi(s)\|_2^2=\int_0^t\|h(s)\|_2^2ds.\]
\end{theorem}
The proof \cite{Barnett2} is similar to Ito's.
We use this to construct the integral of square-integrable adapted processes,
and some $L^p$ processes, by extension to the completion of the space of
simple adapted processes. The stochastic integral is the quantised field
$\Psi$, smeared with an operator $h$ rather than a test-function.
There is a Doob-Meyer theorem: $M_t^2$ is the sum of a martingale, denoted by
$[M_t,M_t]$ in classical theory, (NOT the commutator!) and an increasing
process of bounded variation, denoted $\langle M_t,M_t\rangle$.
Any stochastic integral is a martingale, and
we show the converse, that any $L^2$ martingale of mean zero is a
stochastic integral.
We also define the stochastic integral $N(t)=\int_0^th(s)
dM(s)$, where $h$ is adapted and square-integrable relative to $\langle M_t,
M_t\rangle$. Here,
$M$ is an $L^2$-martingale. This representation of
$N$ is unique; we then write $h$ as the stochastic derivative: $h=\partial
N/\partial M$. We show that we can change variables in the
integral: the stochastic Radon-Nikodym theorem \cite{Barnett}.

We are able to show \cite{Barnett5} that the quantum sde
\begin{equation}
dX_t=F(X_t,t)dM_t+dM_tG(X_t,t)+H(X_t,t)dt
\end{equation}
has a solution in $L^2({\cal A},\varphi)$ for $F,G,H$ continuous, adapted
and locally uniformly Lipschitz, for any martingale $M_t$ of degree $n$,
and that the solution obeys the Markov property \cite{Barnett6}.

Manipulations of differentials
are similar to the Ito calculus: $(dt)^2=0=(dt)(d\Psi)$; $(d\Psi)^2=dt$.
Pisier and Xu have obtained `Burkholder-Gundy' inequalities
within this theory \cite{Pisier}.

The central state $\varphi$ of the Clifford algebra corresponds physically
to an infinite temperature.
For the {\em CCR} and {\em CAR} algebras, we constructed the stochastic
integrals starting with quasifree states with no Fock part, using the
non-central state in place of the trace \cite{Barnett4,Lindsay}. This theory
is somewhat technical (`unreadable' \cite{Meyer2}).

The general Lindblad semigroup can be dilated \cite{Evans}
using the flow defined by a solution to a quantum stochastic equation in
the sense of Hudson and Parthasarathy \cite{Hudson3,Partha2,Smith}.
It was extended to some unbounded cases by Belavkin \cite{Belavkin}.
We now give a brief account of this, following Frigerio \cite{Frigerio}.

Let $T_t=\exp({\cal L}t)$ be a semigroup of completely positive normal
stochastic maps on the algebra ${\cal B}({\cal H})$.
\begin{theorem}
There exists a Hilbert space ${\cal F}$, a group $\{\alpha_t:t\in{\bf R}\}$
of $^*$-automorphisms of ${\cal B}({\cal H}\otimes{\cal F})$ and a
conditional expectation $E_0$ of ${\cal B}({\cal H}\otimes{\cal F})$
onto ${\cal B}({\cal H})\otimes I_{\cal F}$ such that
\begin{equation}
T_t(X)\otimes I_{\cal F}=E_0[\alpha_t(X\otimes I_{\cal F}0],\hspace{.3in}
X\in{\cal B}({\cal H}),\;\;t\in{\bf R}.
\end{equation}
\end{theorem}
The evolution $\alpha_t$ is a perturbation of the `free evolution'
$\alpha_t^0$ on ${\cal B}({\cal H})$, of the form
\begin{equation}
\alpha_t(\,.\,)=U(t)\alpha_t^0(\,.\,)U(t)^*,
\end{equation}
where $\{U(t):t\in{\bf R}\}$ satisfies the cocycle condition
\begin{equation}
U(t)\alpha_t^0(U(s))=U(s+t),\hspace{.3in}t,s\in{\bf r},
\end{equation}
is unitary and is the solution of a qsde in the sense of
\cite{Hudson3,Partha2}. We give the details in the simplest case,
eq.~(\ref{Lindblad}) with only one term $S$ in the sum. We take ${\cal F}=
\Gamma(L^2{\bf R})$, with total the set of coherent vectors $\exp\phi:\phi
\in L^2({\bf R})\cap L^1({\bf R})$. We define the {\em annihilation} process,
{\em creation} process and {\em gauge} process on this total set by
\begin{eqnarray}
A(t)\exp\phi&=&(\int_0^t\phi(s)ds)\exp\phi\\
A^*(t)\exp\phi&=&\frac{d}{d\epsilon}\exp\left(\phi+\epsilon\chi_{[0,t]}
\right)|_{\epsilon=0}\\
\Lambda(t)\exp\phi&=&\frac{d}{d\epsilon}\exp\left(e^{\epsilon\chi_{[0,t]}}
\phi\right)|_{\epsilon=0}.
\end{eqnarray}
The conditional expectation $M_t$ is as for the {\em CTP},
$\otimes_{s=0}^t({\cal F}_s)$, based on the Fock vacuum, and
$A(t),A^*(t)$ are the creators and annihilators defined by
the generators $P,Q$ of the Heisenberg
subgroup of the oscillator group; $\Lambda$ is the number operator.

We identify any operator $X$ in ${\cal B}({\cal H})$ with its ampliation
$X\otimes I_{\cal F}$, and any operator $Y$ with domain ${\cal D}
\subseteq{\cal F}$ with the algebraic tensor product $I_{\cal H}\otimes Y$.
A family $U(t)$ is found by solving the qsde
\begin{equation}
dU(t)=U(t)\left[iS^*dA(t)+iS\,dA^*(t)+(iH-S^*S/2)dt\right],
\end{equation}
with the initial condition $U(0)=I$. The structure of the equation is
designed to ensure that the solution, defined on the set of coherent states,
is continuous, unitary, and adapted. The term $S^*S/2$ arises as the Ito
correction, or as due to the Wick ordering \cite{Hudson2}.
To ensure that $\alpha_t$ obeys the group law, the usual free evolution
$\alpha^0$ on ${\cal F}$, the second quantisation of the
translation group on $L^2({\bf R})$, is chosen. It is then proved that
$\alpha^0_{-s}[U^*(s)U(s+t)]$ satisfies the same qsde as $U(t)$, and so,
by uniqueness, must be $U(t)$. So $U$ satisfies the cocycle condition.
On multiplying out, we see that $\alpha_y$ is a group.

The theorem for a semigroup with a finite number of operators $S_j$
follows a similar line.\hspace{\fill}$\Box$\\
There is a fermionic version of this dilation \cite{Applebaum}.

Quantum stochastic calculus has become a mature field of mathematics.
The approach of \cite{Hudson3}, rather than \cite{Barnett2}, has the
disadvantage that the stochastic integrals are defined as operators
only on a dense set. It is not always clear that they have a unique closed
extension.
This is overcome in \cite{Hudson3,Partha2} by limiting the class of
equations to those with unitary solutions.  Another help in the analysis
is by the use of Maassen kernels \cite{Maassen}. Alternatively, \cite{Obata}
one may give a meaning to these objects as maps between test-functions and
distributions, using white-noise analysis.

One problem with this work, and this includes \cite{Barnett}
as well, is that the spectrum of the noise is white, so that random
negative energy is added as well as positive energy.
We saw that positive energy seems to exclude martingales
\cite{Senitzky,RFS5}. In fact, the {\em KMS} condition excludes the existence
of a conditional expectation except in trivial cases. It has been remarked
that it also excludes the Markov property and the regression theorem
\cite{Talkner}. Lindblad has remarked \cite{Lindblad2} that for the
oscillator, the {\em KMS} condition is not compatible with the axioms of
dynamical semigroups. So to model random external forces in a real system,
coupled to a heat-bath, the white noise sde is an approximation, that
might be good if the time-interval is large compared with the memory time.
These ideas are used
to describe quantum systems like lasers, which are subject to external
forces; this was the original intention of Senitzky and Lax. The
modern version is described in \cite{Alicki}. Since external forces
introduce energy and entropy into a system, such models have two
drawbacks:
\begin{enumerate}
\item The first law of thermodynamics is not obeyed.
\item The second law of thermodynamics is not obeyed.
\end{enumerate}
This is the starting point of \cite{AlickiM,Balian,Streater}.
One step of the linear dynamics is given by a bistochastic map
$\rho\mapsto\rho T$, so that entropy increases. We require that $T^*$ maps
any spectral projection of the energy to itself; this will preserve energy.
To reduce the description, we then project the
new state $\rho T$ onto the information manifold ${\cal M}$ defined by
the set of slow variables, to get the state $\rho TQ$.
To preserve mean energy, the energy must be a slow variable. The map $Q$
is nonlinear and is interpreted as the thermalisation of the fast
variables. Thus, after the map $T$, the system itself decides to find the
best estimate $\rho TQ$ to $\rho T$ within ${\cal M}$. The resulting map
gives a nonlinear motion through the manifold, obeying the first and second
laws of thermodynamics. This theory, called {\em statistical dynamics},
is still being explored \cite{RFS1,RFS7,Streater}.

\end{document}